\documentclass{ws-mpla}
\usepackage[super]{cite}
\usepackage{amsmath,amssymb,graphics,epsfig,graphicx,diagbox,slashbox,multirow,floatrow,subfigure,color,hyperref}
\begin{document}
\markboth{Nan Zhang}
{Diagnosing Tsallis Holographic Dark Energy models with interactions}

\catchline{}{}{}{}{}

\title{Diagnosing Tsallis Holographic Dark Energy models with interactions}

\author{\footnotesize Nan Zhang\footnote{zhangnandalian@163.com}}
\author{\footnotesize Ya-Bo Wu\footnote{Corresponding author: ybwu61@163.com}}
\author{Jia-Nan Chi}
\author{Zhe Yu}
\author{Dong-Fang Xu}

\address{Department of Physics, Liaoning Normal University, Dalian 116029, P.R.China}

\maketitle

\pub{Received (Day Month Year)}{Revised (Day Month Year)}

\begin{abstract}
It has been found that the geometrical diagnostic methods can break the degeneracy for dark energy models.
In this paper, we investigate the $Om$ diagnostic, the statefinder hierarchy $S_{n}$ and the composite null diagnostic $\{S_{n},\epsilon\}$ for the Tsallis holographic dark energy models with interactions.
We find that model parameters and the forms of interaction will influence the values of diagnostic parameters or the trends of the evolutionary trajectories for each model.
Moreover, the statefinder hierarchy $S_{3}^{(1)}$ together with $\{S_{3}^{(1)},\epsilon\}$ could give good diagnostic results.
Furthermore, we also obtain some issues of cosmological structure by means of the composite null diagnostic.

\end{abstract}

\ccode{PACS Nos.: 98.80.-k, 95.36.+x, 95.35.+d, 98.80.Es.}

\section{Introduction}	

The observation results imply that the expansion of our current universe is accelerating \cite{Riess:1998cb,Perlmutter:1998np,Ade:2015xua,Aghanim:2018eyx}.
In order to explain the accelerated expansion in the framework of the standard cosmology, the dark energy (DE) is introduced as an exotic component with negative pressure.
However, the nature of DE still remains mysterious, many kinds of DE models therefore have been constructed \cite{Copeland:2006wr,Bamba:2012cp}.
The simplest one is the cosmological constant model, i.e., $\Lambda$CDM model \cite{Sahni:1999gb}, its energy density is one constant and the equation of state (EoS) is $w_{\Lambda}=-1$.
Although the $\Lambda$CDM model fits good to the observational data, it faces challenges of the fine tuning problem and the coincidence problem.
Thus, the dynamical DE models have been proposed as the alternatives, such as quintessence \cite{Caldwell:1997ii}, phantom \cite{Caldwell:1999ew,Carroll:2003st}, the Chaplygin gas (CG) model \cite{Kamenshchik:2001cp}, the holographic dark energy (HDE) model \cite{Li:2004rb} and the agegraphic dark energy (ADE) model \cite{Cai:2007us} etc.
As we know, HDE models are proposed based on the holographic principle and the systemic entropy, the more details about HDE models can see Ref.~\cite{Wang:2016och}.
In the standard HDE models, the energy density of DE is defined as $\rho_{D}=3c^{2}m_{pl}^{2}L^{-2}$, where the constant $c$ is the model parameter, $L$ represents the IR cutoff, $m_{pl}^{2}$ is the reduced Planck mass.
While recently, considering the long-range character of gravity and the important role of Tsallis generalized entropy in studying the gravitational systems through the generalized statistical mechanics  \cite{Tsallis:2012js,Nunes:2014jra,Nunes:2015xsa}, the authors of Ref.~\cite{Tavayef:2018xwx} adopt the Tsallis entropy instead of the Bekenstein entropy to construct a new class of DE models, namely, Tsallis Holographic Dark Energy (THDE) models \cite{Tavayef:2018xwx}. The energy density of THDE model is $\rho_{D}=BL^{2\delta-4}$, where $B$ and $\delta$ are the model parameters.
Taking the different IR cutoffs $L$, one can obtain different THDE models \cite{Saridakis:2018unr,Zadeh:2018poj,Zadeh:2018wub}.
Moreover, the related content to THDE models in other gravitational theories and its perturbations have also been researched \cite{Ghaffari:2018rzs,Ghaffari:2018wks,DAgostino:2019wko}.
Besides, it should be noted that the observation today also allows a mutual interaction $Q$ between DE and dark matter (DM), more information about the interacting DE models can be seen in Refs.~\cite{Wang:2016lxa,Cai:2012fq,Yang:2017zjs,Li:2018jiu,Yang:2018euj,Clemson:2011an,Valiviita:2009nu,Chimento:2009hj,Li:2015vla}.
Thus, $Q$ can be embedded in THDE models and the corresponding results to the forms of $Q=3b^{2}H(\rho_{D}+\rho_{m})$,  $Q=3b^{2}qH(\rho_{D}+\rho_{m})$ have been discussed in Refs.~\cite{Zadeh:2018poj,Ghaffari:2018wks} etc.

In fact, there are various DE models, interaction forms and model parameters, how to distinguish them becomes an interesting item.
And the differences between the standard $\Lambda$CDM model and other DE models are always the focus because the observation analyses today are mostly based on $\Lambda$CDM model.
Thus, the model-independent geometrical diagnostic methods for DE models have been widely researched, such as the $Om$ diagnostic \cite{Sahni:2008xx,Zunckel:2008ti}, the statefinder diagnostic $\{r, s\}$ \cite{Sahni:2002fz,Alam:2003sc}, and the statefinder hierarchy $S_{n}$ \cite{Arabsalmani:2011fz}.
The $Om$ diagnostic and statefinders are related to the expansion rate $H(z)$ and the derivative of the scale factor $a(t)$, respectively.
In addition, the composite null diagnostic (CND) $\{S_{n},\epsilon\}$ is proposed in Ref.~\cite{Arabsalmani:2011fz} as a useful supplementary method to the statefinder hierarchy, where $\epsilon$ \cite{Acquaviva:2008qp,Acquaviva:2010vr} represents the fractional growth parameter related to the growth rate of structure $f$ \cite{Wang:1998gt,Linder:2005in,CalderaCabral:2009ja,Nodal:2009ny,Wu:2013elu,Sen:2005nr}.

By applying the diagnostic methods to DE models, it has been found that the statefinder hierarchy and CND can break the degeneracy for some class of HDE models \cite{Zhang:2014sqa,Yin:2015pqa,Zhao:2017zcj}.
However, it could not be believed that the higher order of the statefinder, the more effective it is \cite{Yu:2015sla}.
Thus, we will further here investigate THDE models with different interaction forms and perform the geometrical diagnostic methods to the models.
The cutoffs we have taken in this paper include the hubble horizon, the future event horizon and the GO (Granda and Oliveros) horizon \cite{Granda:2008dk,Granda:2008tm,Nojiri:2005pu}, here we call them THDE-H, THDE-f and THDE-GO models, respectively.
The three forms of interaction are adopted as $Q_{1}=3H\xi\rho_{D}$, $Q_{2}=3H\xi\frac{\rho_{m}\rho_{D}}{\rho_{m}+\rho_{D}}$ and $Q_{3}=3H\xi\frac{\rho_{D}^{2}}{\rho_{m}+\rho_{D}}$.
Although the diagnostic for THDE models has been discussed in Refs.~\cite{Sharma:2019mtn,Varshney:2019fzj}, they only consider the related content of the statefinder diagnostic and $w-w'$ pair for THDE-H model with $Q=0$ and $Q=3b^{2}H(\rho_{D}+\rho_{m})$.
But the $Om$ diagnostic, the statefinder hierarchy and CND have not been researched for THDE models with other cutoffs and different forms of $Q$, which is our purpose in this paper.

Based on the above motivations, we will apply the $Om$ diagnostic, the statefinder hierarchy and CND to the three different THDE models in order to explore their geometrical evolutions and structural information. We will also investigate the effectiveness of the different diagnostic method. Of course, the standard $\Lambda$CDM model will be given for comparison. Our research results indicate that the model parameters and forms of interaction will influence the values of diagnostic parameters and the trends of evolutionary curves for the models.

This paper is organized as follows. In Sec.~\ref{brife thde} and Sec.~\ref{brife diagnostic}, we briefly review THDE models with different cutoffs and the common diagnostic methods for DE models, respectively.
In Sec.~\ref{apply to thde}, we apply the diagnostic methods to THDE models in order to discuss the evolutionary trajectories and the diagnostic effectiveness in different cases.
The conclusion is given in Sec.~\ref{conclusion}.

\section{THDE models with different cutoffs}
\label{brife thde}
Considering the density $\rho$ of the Universe is comprised of two components, i.e., $\rho_{D}$ and $\rho_{m}$, thus the total density and the Friedmann equation can be respectively written as $\rho=\rho_{D}+\rho_{m}$ and $H^{2}=\frac{8\pi G}{3}(\rho_{D}+\rho_{m})$.
Define the dimensionless density parameters as follows:
\begin{equation}\label{omod}
\Omega_{D}=\frac{\rho_{D}}{3m_{p}^{2}H^{2}}~~and~~\Omega_{m}=\frac{\rho_{m}}{3m_{p}^{2}H^{2}}\mbox{,}
\end{equation}
where $\Omega_{D}+\Omega_{m}=1$. The continuity equation of DE and DM can be expressed as:
\begin{equation}\label{cmcd}
\dot{\rho_{D}}+3H(1+w_{D})\rho_{D}=-Q~~and~~\dot{\rho_{m}}+3H\rho_{m}=Q\mbox{,}
\end{equation}
where the dot denotes differentiation with respect to cosmic time $t$, $w_{D}$ is the equation of state of DE, and $w_{m}=0$ for DM has been used in the above equation.
$Q$ is the energy transfer between DE and DM, and it's obvious that there is no interaction between DE and DM when $Q=0$.
Besides, the total density is always conserved, i.e., $\dot{\rho}+3H(1+w)\rho=0$.
Taking the time derivative of the Friedmann equation and making use of Eqs.~(\ref{omod}), (\ref{cmcd}), one can obtain the following equation:
\begin{equation}\label{hw}
\frac{\dot{H}}{H^{2}}=-\frac{3}{2}(1+w_{D}\Omega_{D})\mbox{.}
\end{equation}

Considering the holographic hypothesis and the general Tsallis's entropy expression \cite{Tsallis:2012js}, i.e., $S_{\delta}=\gamma A^{\delta}$, following the relation between the system entropy (S),the IR (L) and UV ($\Lambda$), the density $\rho_{D}$ for THDE models can finally be written as $\rho_{D}=BL^{2\delta-4}$ \cite{Tavayef:2018xwx}, where $B, \delta$ are the model parameters. It's obvious that the density of the standard HDE model $\rho_{D}=3c^{2}m_{pl}^{2}L^{-2}$ \cite{Li:2004rb} can be obtained at the appropriate limits.
Besides, we chose the relatively general form of the interaction term $Q$ as $Q=3\xi H\rho_{m}^{\lambda}\rho_{D}^{1-\lambda-\gamma}(\rho_{m}+\rho_{D})^{\gamma}$ \cite{Yang:2017zjs}, where $\xi$ represents the interaction strength. As we can see, the usual form $Q=3H\xi\rho_{D}$ corresponds to the case of $\lambda=0, \gamma=0$; $Q=3H\xi\rho_{m}$ corresponds to $\lambda=1, \gamma=0$; $Q=3H\xi(\rho_{m}+\rho_{D})$ corresponds to $\lambda=0, \gamma=1$ \cite{Clemson:2011an}-\cite{Chimento:2009hj}.

\subsection{THDE-H model}
Taking the Hubble horizon as the IR cutoff, i.e., $L=H^{-1}$, the density of DE in THDE-H model can be written as \cite{Tavayef:2018xwx}
\begin{equation}\label{hr}
\rho_{D}=BH^{-2\delta+4}\mbox{.}
\end{equation}
By substituting Eq.~(\ref{hr}) into Eq.~(\ref{cmcd}), we can obtain
\begin{eqnarray}\label{hwd}
w_{D}=\frac{\delta-1+\xi\rho_{m}^{\lambda}\rho_{D}^{-\lambda-\gamma}(\rho_{m}+\rho_{D})^{\gamma}}{(2-\delta)\Omega_{D}-1}\mbox{.}
\end{eqnarray}
Based on Eqs.~(\ref{hw}) and (\ref{hwd}), one can deduce that
\begin{eqnarray}\label{hhw}
\frac{\dot{H}}{H^{2}}=-\frac{3}{2}\frac{\Omega_{D}-1+\xi\Omega_{D}\rho_{m}^{\lambda}\rho_{D}^{-\lambda-\gamma}(\rho_{m}+\rho_{D})^{\gamma}}{(2-\delta)\Omega_{D}-1}\mbox{.}
\end{eqnarray}
From Eq.~(\ref{hr}), we can get $\Omega_{D}=\frac{B}{3m_{p}^{2}}H^{-2\delta+2}$, thus we have
\begin{equation}\label{hoddn}
\Omega_{D}'=\frac{d\Omega_{D}}{d\ln a}=\frac{\dot{\Omega_{D}}}{H}=3(\delta-1)\Omega_{D}\frac{1-\Omega_{D}-\xi\Omega_{D}\rho_{m}^{\lambda}\rho_{D}^{-\lambda-\gamma}(\rho_{m}+\rho_{D})^{\gamma}}{1-(2-\delta)\Omega_{D}}\mbox{.}
\end{equation}

\subsection{THDE-f model}
Taking the future event horizon as the IR cutoff, i.e., $L=R_{h}$, the density of the DE in THDE-f model can be written as \cite{Saridakis:2018unr}
\begin{equation}\label{fr}
\rho_{D}=BR_{h}^{2\delta-4}\mbox{,}
\end{equation}
where $R_{h}$ is the future event horizon, defined as $R_{h}\equiv a\int_{t}^{\infty}\frac{dt}{a}$.
It gives that $\dot{R_{h}}=HR_{h}-1$.
Besides, from Eqs.~(\ref{omod}) and (\ref{fr}), we can obtain the relation $R_{h}=(\frac{3m_{p}^{2}H^{2}\Omega_{D}}{B})^{\frac{1}{2\delta-4}}$.
Thus, the time differentiation of $\rho_{D}$ can be expressed as
\begin{equation}\label{frdt}
\dot{\rho_{D}}=(2\delta-4)\rho_{D}H[1-(\frac{3m_{p}^{2}H^{2\delta-2}\Omega_{D}}{B})^{\frac{1}{4-2\delta}}]\mbox{,}
\end{equation}
Substituting Eq.~(\ref{frdt}) into Eq.~(\ref{cmcd}), we can get the parameter $w_{D}$ as
\begin{equation}\label{fwd}
w_{D}=-1-\frac{2\delta-4}{3}[1-(\frac{3m_{p}^{2}H^{2\delta-2}\Omega_{D}}{B})^{\frac{1}{4-2\delta}}]-\xi\rho_{m}^{\lambda}\rho_{D}^{-\lambda-\gamma}(\rho_{m}+\rho_{D})^{\gamma}\mbox{.}
\end{equation}
Taking the time differentiation of $\Omega_{D}$, we finally get
\begin{eqnarray}\label{fodf}
\Omega_{D}'=\frac{\dot{\Omega_{D}}}{H}=\Omega_{D}(1-\Omega_{D})[2(\frac{3m_{p}^{2}H^{2\delta-2}\Omega_{D}}{B})^{\frac{1}{4-2\delta}}(2-\delta)+2\delta-1]-3\xi\Omega_{D}^{2}\rho_{m}^{\lambda}\rho_{D}^{-\lambda-\gamma}(\rho_{m}+\rho_{D})^{\gamma}\mbox{.}
\end{eqnarray}

\subsection{THDE-GO model}
Granda and Oliveros (GO) presented a cutoff to solve the causality and coincidence problems, i.e., GO cutoff, defined as $L=(\alpha H^{2}+\beta\dot{H})^{-1/2}$ \cite{Granda:2008dk,Granda:2008tm}.
Thus, the density of DE in THDE-GO model can be written as \cite{Zadeh:2018poj}
\begin{equation}\label{gor}
\rho_{D}=B(\alpha H^{2}+\beta \dot{H})^{2-\delta}\mbox{,}
\end{equation}
where $\alpha$, $\beta$ are constants. Thus, we have
\begin{equation}\label{goh}
\frac{\dot{H}}{H^{2}}=\frac{1}{\beta}[\frac{(\frac{3m_{p}^{2}\Omega_{D}}{B})^{\frac{1}{2-\delta}}}{H^{\frac{2-2\delta}{2-\delta}}}-\alpha]\mbox{.}
\end{equation}
Similarly, taking the time differentiation of $\Omega_{D}$, using the Friedmann equation and Eq. (\ref{cmcd}), we can get
\begin{eqnarray}\label{good2}
\Omega_{D}'=(1-\Omega_{D})[\frac{2}{\beta}(\frac{(\frac{3m_{p}^{2}\Omega_{D}}{B})^{\frac{1}{2-\delta}}}{H^{\frac{2-2\delta}{2-\delta}}}-\alpha)+3]-3\xi\rho_{m}^{\lambda}\rho_{D}^{1-\lambda-\gamma}(\rho_{m}+\rho_{D})^{\gamma-1}\mbox{.}
\end{eqnarray}
Making use of Eqs.~(\ref{goh}) and (\ref{good2}), a set of solutions $\{\Omega_{D}, H\}$ can be obtained.
On the other hand, considering Eq.~(\ref{hw}), the parameter $w_{D}$ can be expressed as follows
\begin{equation}\label{gowd}
w_{D}=-\frac{1}{\Omega_{D}}-\frac{2}{3\beta\Omega_{D}}[\frac{(\frac{3m_{p}^{2}\Omega_{D}}{B})^{\frac{1}{2-\delta}}}{H^{\frac{2-2\delta}{2-\delta}}}-\alpha]\mbox{.}
\end{equation}

\section{The diagnostic methods}
\label{brife diagnostic}

Firstly, the $Om$ diagnostic is defined as \cite{Sahni:2008xx,Zunckel:2008ti}
\begin{equation}\label{om}
Om(x)=\frac{h^{2}(x)-1}{x^{3}-1},~~~~~~x\equiv1+z\mbox{,}
\end{equation}
where $h(x)=\frac{H(x)}{H_{0}}$.
The $Om$ diagnostic provides a null test of the $\Lambda$CDM model, i.e., $Om(x)-\Omega_{m}^{0}=0$.
Thus, it can be used as the diagnostic method for diagnosing DE models.

Secondly, the statefinder hierarchy is based on the scale factor $a$, $a(t)/a_{0}=(1+z)^{-1}$ can be expanded around the present epoch $t_{0}$ as follows:
\begin{equation}\label{at}
\frac{a(t)}{a_{0}}=1+\sum_{n=1}^{\infty} \frac{A_{n}(t_{0})}{n!}[H_{0}(t-t_{0})]^{n}, ~~~A_{n}=\frac{a(t)^{(n)}}{a(t)H^{n}}\mbox{,}
\end{equation}
where $a(t)^{(n)}=d^{n}a(t)/dt^{n}$ and $n$ represents a positive integer.
The statefinder hierarchy $S_{n}$ is defined as follows \cite{Arabsalmani:2011fz}:
\begin{equation}\label{s}
S_{2}=A_{2}+\frac{3}{2}\Omega_{m},~~S_{3}=A_{3},~~and ~~S_{4}=A_{4}+\frac{9}{2}\Omega_{m}\mbox{.}
\end{equation}
These equations provide a series of diagnostics for $\Lambda$CDM model with $n\geq3$, i.e., $S_{n}|\Lambda CDM=1$.
Making use of the relation $\Omega_{m}=\frac{2}{3}(1+q)$ for $\Lambda$CDM model, the statefinder hierarchy $S_{3}^{(1)}$, $S_{4}^{(1)}$ can be rewritten as follows:
\begin{equation}\label{sa}
S_{3}^{(1)}=A_{3}~~and~~S_{4}^{(1)}=A_{4}+3(1+q)\mbox{.}
\end{equation}
For $\Lambda$CDM model, $S_{n}^{(1)}=1$.
According to Ref.~\cite{Yin:2015pqa}, for the dynamical dark energy models with interaction term $Q=3\xi H\rho_{m}^{\lambda}\rho_{D}^{1-\lambda-\gamma}(\rho_{m}+\rho_{D})^{\gamma}$ \cite{Yang:2017zjs} between DE and DM, the expressions of $S_{3}^{(1)}$ and $S_{4}^{(1)}$ can be deduced as follows:
\begin{equation}\label{s31f}
S_{3}^{(1)}=1+\frac{9}{2}w_{D}\Omega_{D}(1+w_{D}\Omega_{D})-\frac{3}{2}w_{D}'\Omega_{D}
+\frac{9}{2}w_{D}\xi\rho_{m}^{\lambda}\rho_{D}^{1-\lambda-\gamma}(\rho_{m}+\rho_{D})^{\gamma-1}\mbox{,}
\end{equation}
\begin{eqnarray}\label{s41f}
S_{4}^{(1)}&=&1-\frac{9}{4}w_{D}\Omega_{D}^{2}[3w_{D}(1+w_{D})-w_{D}']-\frac{3}{4}[w_{D}(21+39w_{D}+18w_{D}^{2})-(13+18w_{D})w_{D}'+2w_{D}'']\Omega_{D}\nonumber\\
&&-\frac{9}{2}w_{D}\xi\rho_{m}^{\lambda}\rho_{D}^{1-\lambda-\gamma}(\rho_{m}+\rho_{D})^{\gamma-1}(2+3w_{D})+9w_{D}'\xi\rho_{m}^{\lambda}\rho_{D}^{1-\lambda-\gamma}(\rho_{m}+\rho_{D})^{\gamma-1}\nonumber\\
&&+\frac{9}{2}w_{D}\xi[(\rho_{m}^{\lambda}\rho_{D}^{1-\lambda-\gamma}(\rho_{m}+\rho_{D})^{\gamma-1})'-\frac{9}{2}(1+w_{D}\Omega_{D})\rho_{m}^{\lambda}\rho_{D}^{1-\lambda-\gamma}(\rho_{m}+\rho_{D})^{\gamma-1}]\mbox{.}
\end{eqnarray}
If giving the specific values of $\lambda$ and $\gamma$, then we can obtain the corresponding forms of $S_{3}^{(1)}$ and $S_{4}^{(1)}$.

In addition, the Statefinder hierarchy $S_{n}$ together with the fractional growth parameter $\epsilon$ has been used to define a composite null diagnostic (CND) $\{S_{n},\epsilon\}$ \cite{Arabsalmani:2011fz}.
The fractional growth parameter is defined as \cite{Acquaviva:2008qp,Acquaviva:2010vr}
\begin{equation}
\epsilon(z)\equiv\frac{f(z)}{f_{\Lambda CDM}(z)} \mbox{,}
\end{equation}
where $f(z)=\frac{d\ln \delta}{d\ln a}$ is the growth rate of structure and $\delta$ is the matter density contrast.
When $Q=0$, the growth rate $f(z)$ can be approximately expressed as $f(z)\simeq\Omega_{m}(z)^{\gamma}$ and $\gamma(z)\simeq\frac{3}{5-\frac{w_{D}}{1-w_{D}}}+\frac{3}{125}\frac{(1-w_{D})(1-3/2w_{D})}{(1-6/5w_{D})^{3}}$ for the models with constant or slowly varying EoS \cite{Wang:1998gt}.
For $\Lambda$CDM model, $\gamma(z)\simeq0.55$ and $\epsilon(z)=1$ \cite{Wang:1998gt,Linder:2005in}.
However, when $Q\neq0$, the growth rate can not be simply approximated as above.
It should be obtained by solving the linear perturbation equation numerically \cite{CalderaCabral:2009ja,Nodal:2009ny,Wu:2013elu,Yin:2015pqa}.
In this paper, our initial condition is chosen as $f(z_{LSS})=1$, which is the same as the ones in Refs.~\cite{Nodal:2009ny,Wu:2013elu} and $z_{LSS}$ means the redshift of the last scattering surface.
It's obvious that CND can provide both the geometrical information and the matter perturbation information of cosmic evolution.

\section{Diagnosing THDE models}
\label{apply to thde}
Below, we will apply the geometrical diagnostic methods to three THDE models and fix $\Omega_{D}(z=0)=0.73$ \cite{Tavayef:2018xwx,Zadeh:2018poj,Sharma:2019mtn,Varshney:2019fzj,Malekjani:2010nk}, $H(z=0)=67$\cite{Aghanim:2018eyx,Zadeh:2018poj}, the specific values of the other related parameters will be shown in the following figures which are chosen to provide the proper description of the late time acceleration expansion of the universe, and the forms of $Q$ are respectively taken as $Q_{1}=3H\xi\rho_{D}$ ($\lambda=0, \gamma=0$), $Q_{2}=3H\xi\frac{\rho_{m}\rho_{D}}{\rho_{m}+\rho_{D}}$ ($\lambda=1, \gamma=-1$) and $Q_{3}=3H\xi\frac{\rho_{D}^{2}}{\rho_{m}+\rho_{D}}$ ($\lambda=0, \gamma=-1$).

\begin{figure}
\centering
\subfigure[]{\label{fig:omnona}\includegraphics[width=5cm,height=5cm]{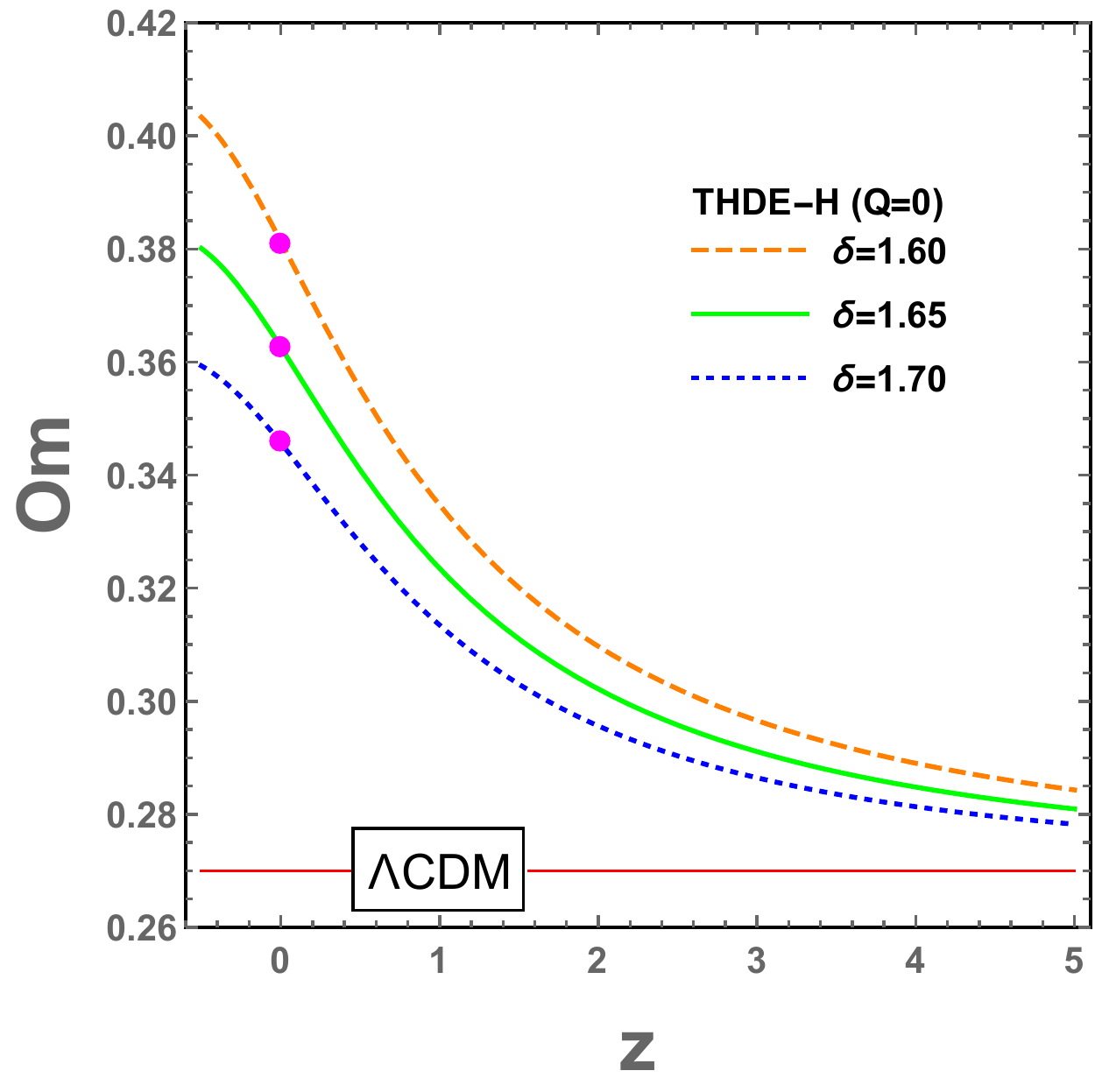}}
\subfigure[]{\label{fig:omnonb}\includegraphics[width=5cm,height=5cm]{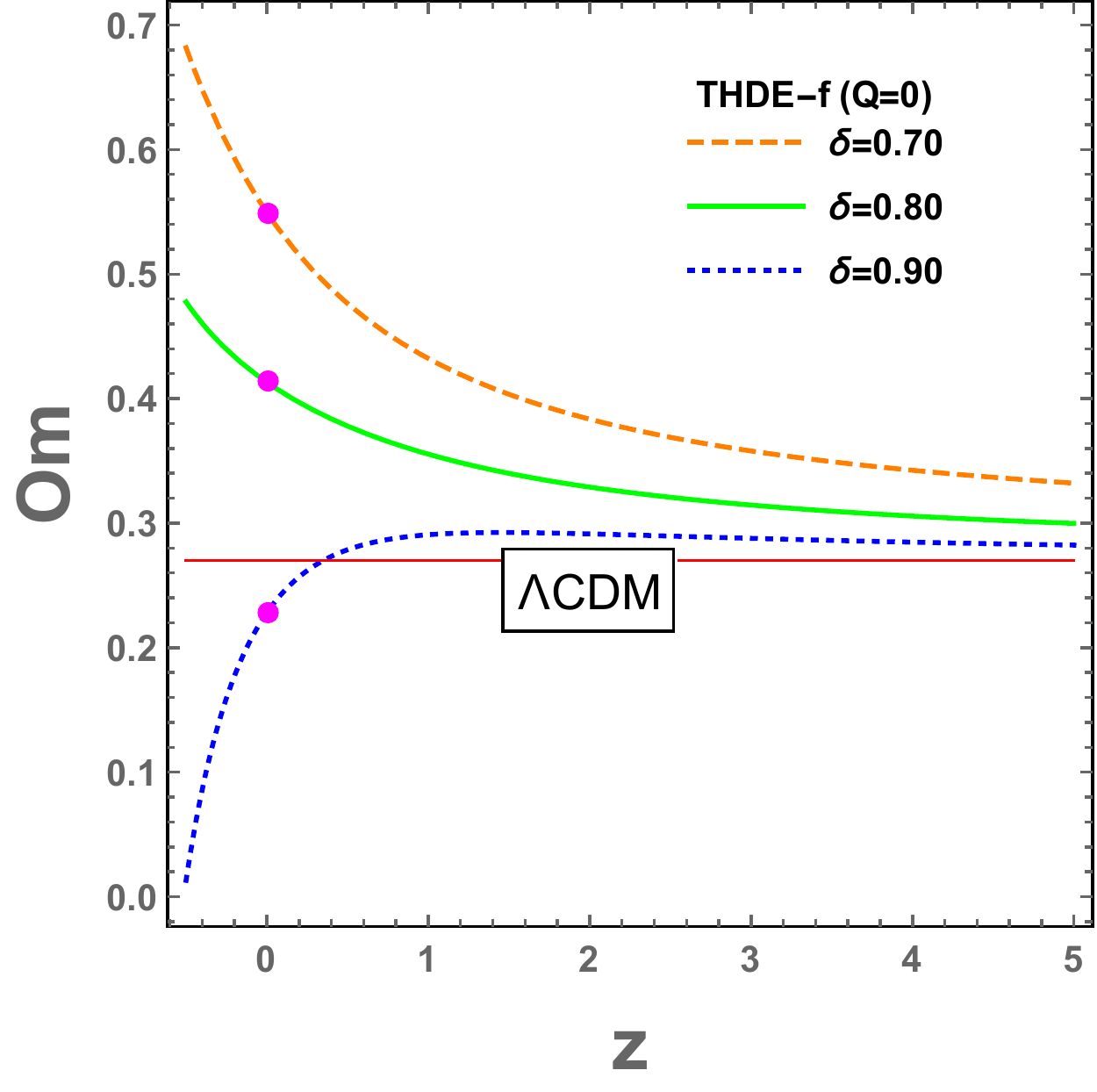}}
\subfigure[]{\label{fig:omnonc}\includegraphics[width=5cm,height=5cm]{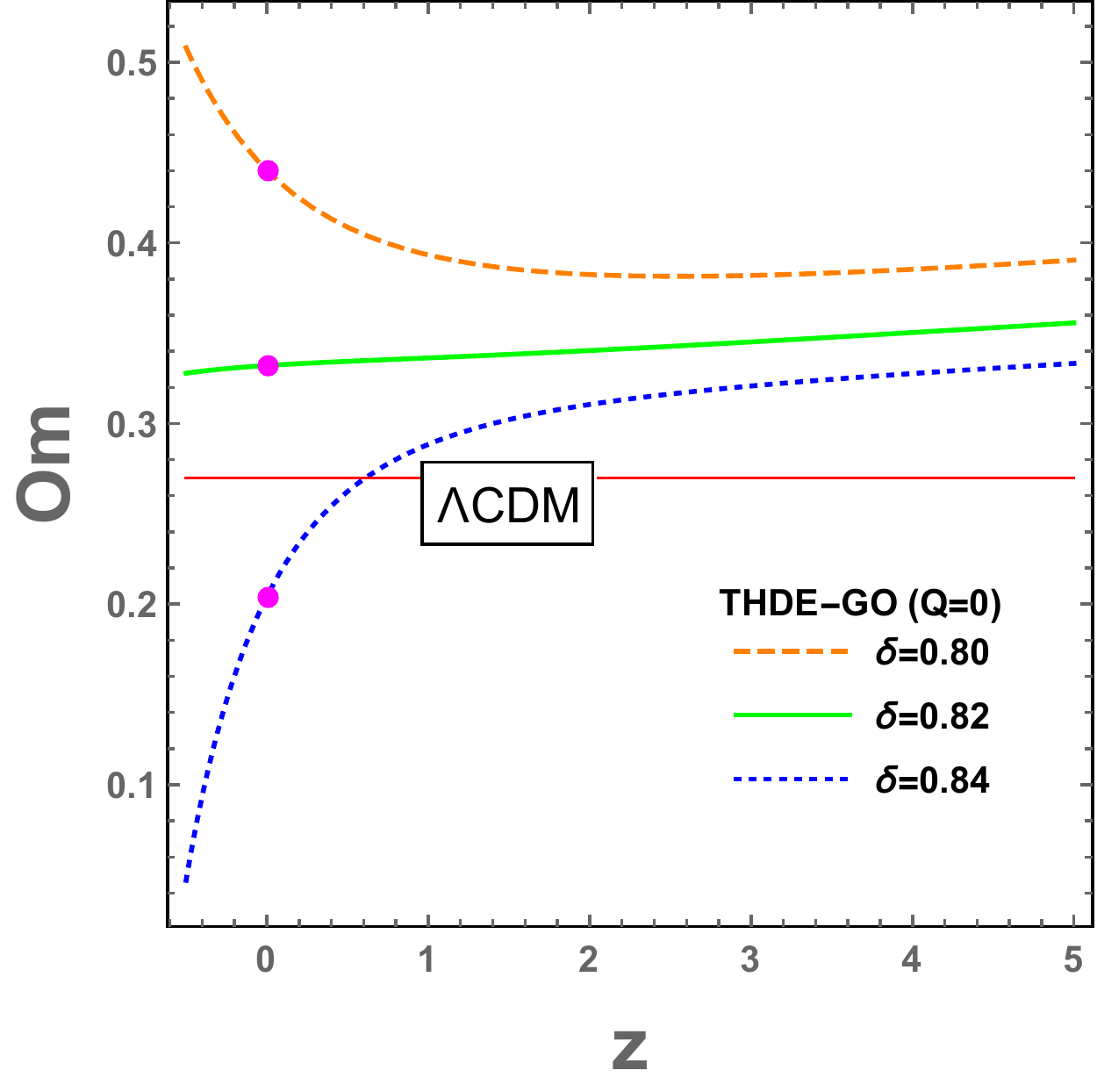}}
\caption{\small{The evolutionary trajectories of $Om(z)$ for THDE models with $Q=0$, where $B=0.8, \alpha=0.9, \beta=0.5.$}}
\label{fig:omnon} 
\end{figure}

\begin{figure}
\centering
\subfigure[]{\label{fig:omqa}\includegraphics[width=5cm,height=5cm]{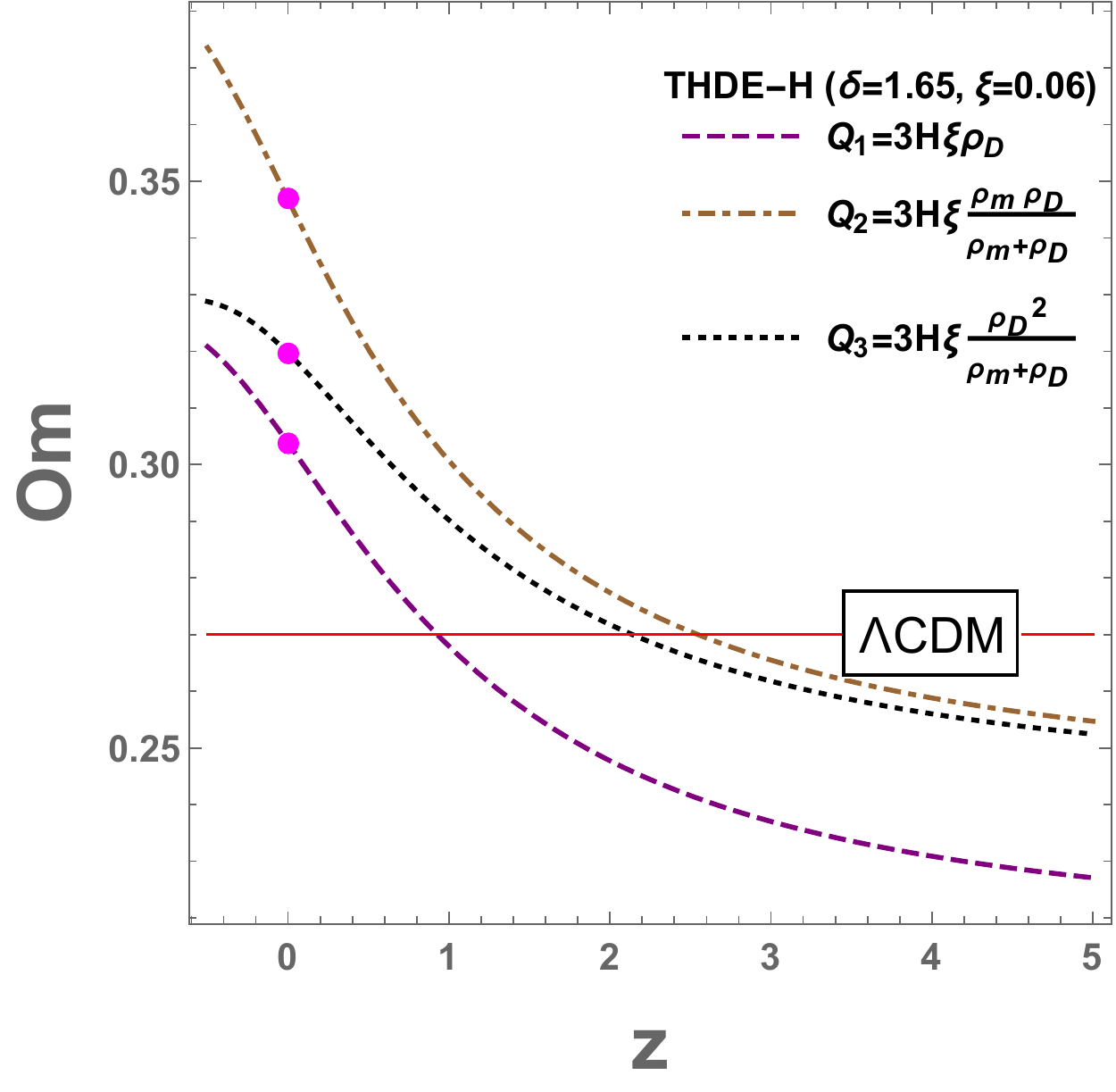}}
\subfigure[]{\label{fig:omqb}\includegraphics[width=5cm,height=5cm]{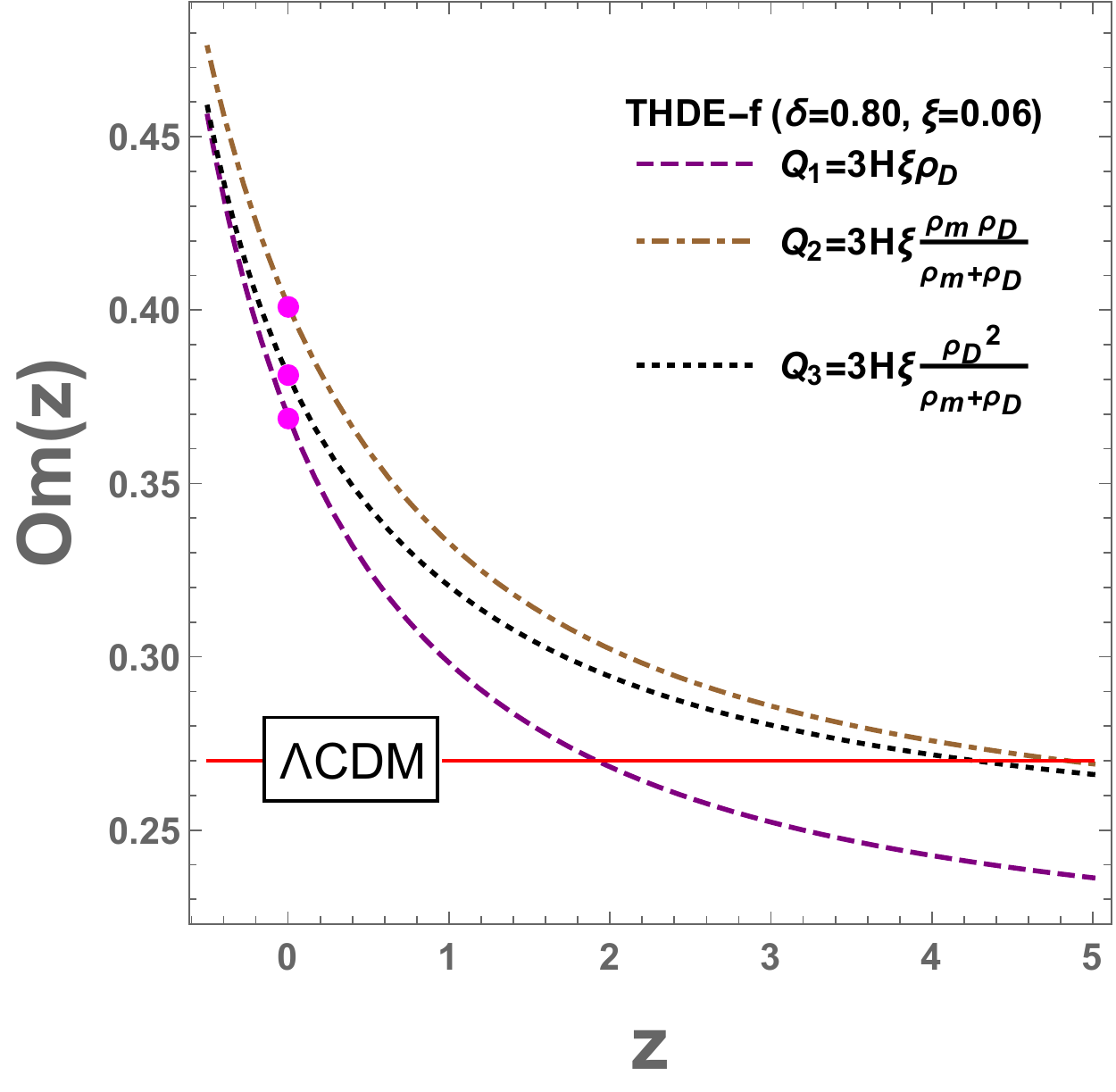}}
\subfigure[]{\label{fig:omqc}\includegraphics[width=5cm,height=5cm]{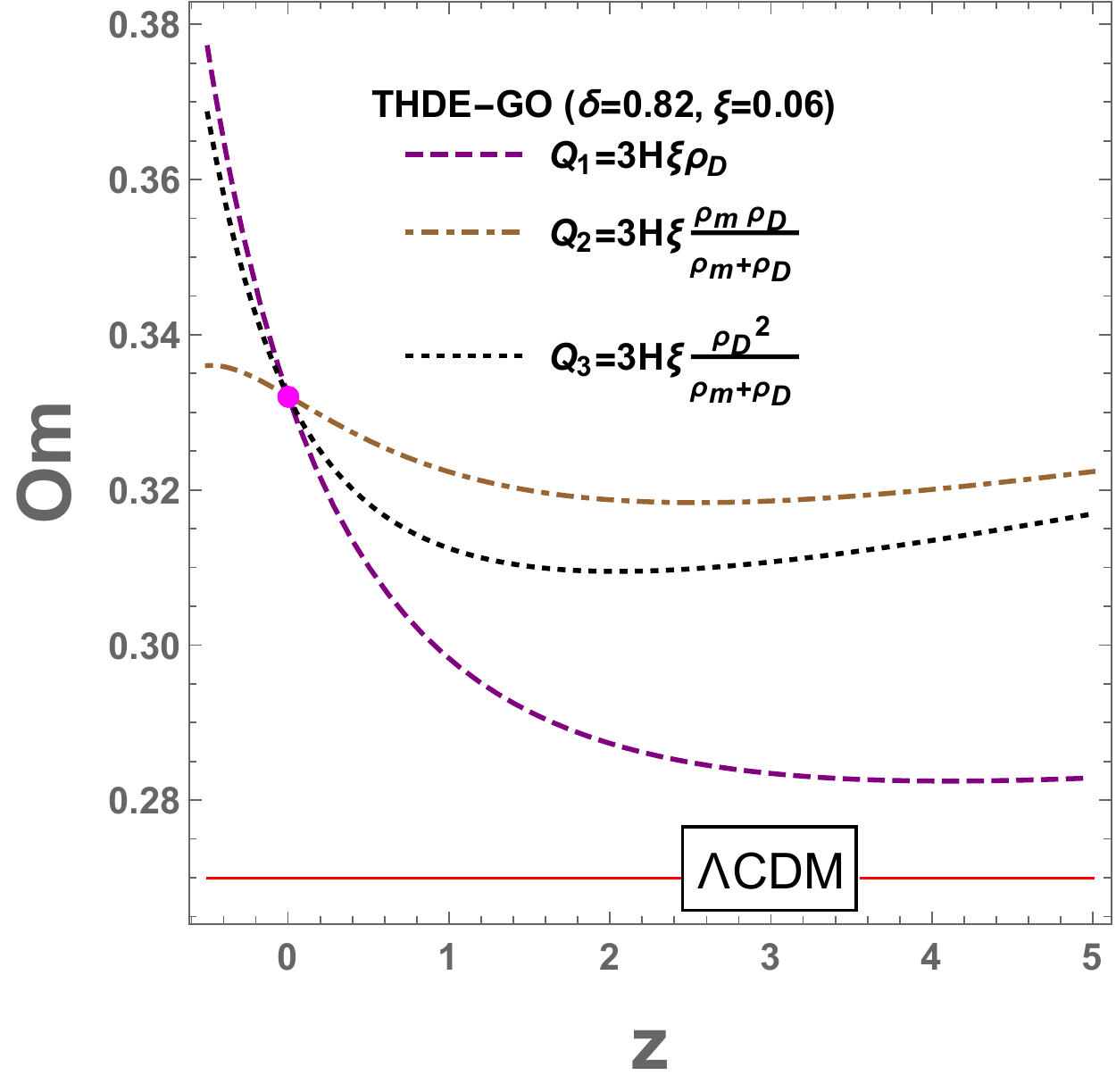}}
\caption{\small{The evolutionary trajectories of $Om(z)$ for interacting THDE models, where $B=0.8, \alpha=0.9, \beta=0.5.$}}
\label{fig:omq} 
\end{figure}

The evolutionary trajectories of the $Om$ versus $z$ for THDE models are respectively plotted in FIG.~\ref{fig:omnon} and FIG.~\ref{fig:omq}.
It can be found from Fig.~\ref{fig:omnon}, the $Om$ diagnostic can efficiently diagnose THDE models when $Q=0$.
It also indicates that taking the different values of model parameter $\delta$ only influences the values of $Om(z)$ for THDE-H model, however, it changes the evolutionary trends of THDE-f and THDE-GO models (see FIGs.~\ref{fig:omnonb}, \ref{fig:omnonc}).
Besides, according to FIG.~\ref{fig:omqa}, one can find that the $Om$ diagnostic does well in diagnosing interacting THDE-H model, while the diagnostic results are unsatisfied for interacting THDE-f and THDE-GO models.
Concretely, for THDE-f model, the corresponding results to $Q_{1}$ and $Q_{3}$ tend to overlap in the future and the curve of $Q_{2}$ is indistinguishable from $\Lambda$CDM model in high-redshift (see FIG.~\ref{fig:omqb}), the present values of $Om(z)$ are almost the same for THDE-GO model in FIG.~\ref{fig:omqc}.
Moreover, comparing the curves in FIG.~\ref{fig:omq} with the corresponding solid curves in FIG.~\ref{fig:omnon}, we find that the interaction $Q$ can significantly influence the starting regions or trends of the evolutionary trajectories for THDE models.

\begin{figure}
\centering
\subfigure[]{\label{fig:s3nona}\includegraphics[width=5cm,height=5cm]{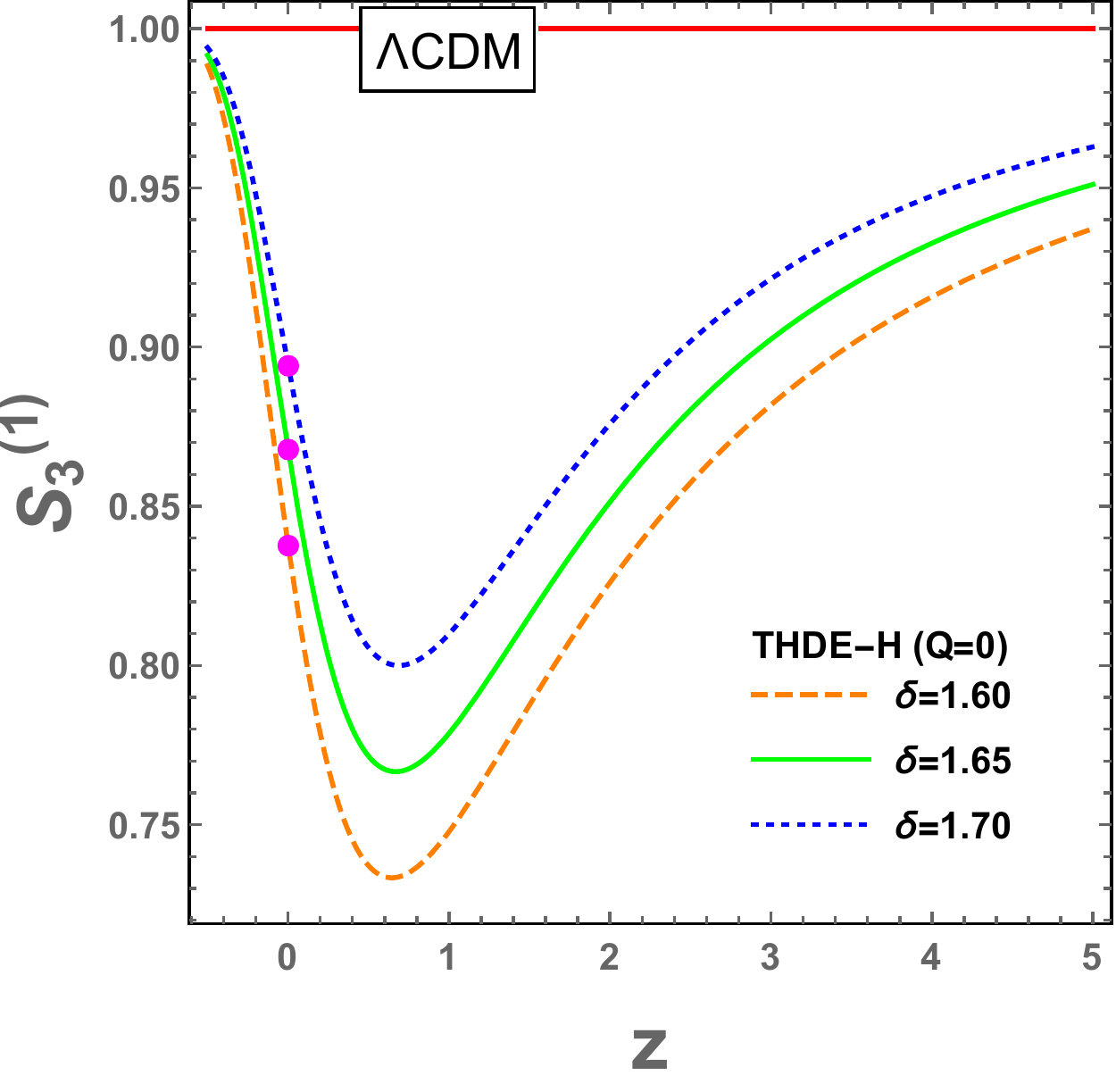}}
\subfigure[]{\label{fig:s3nonb}\includegraphics[width=5cm,height=5cm]{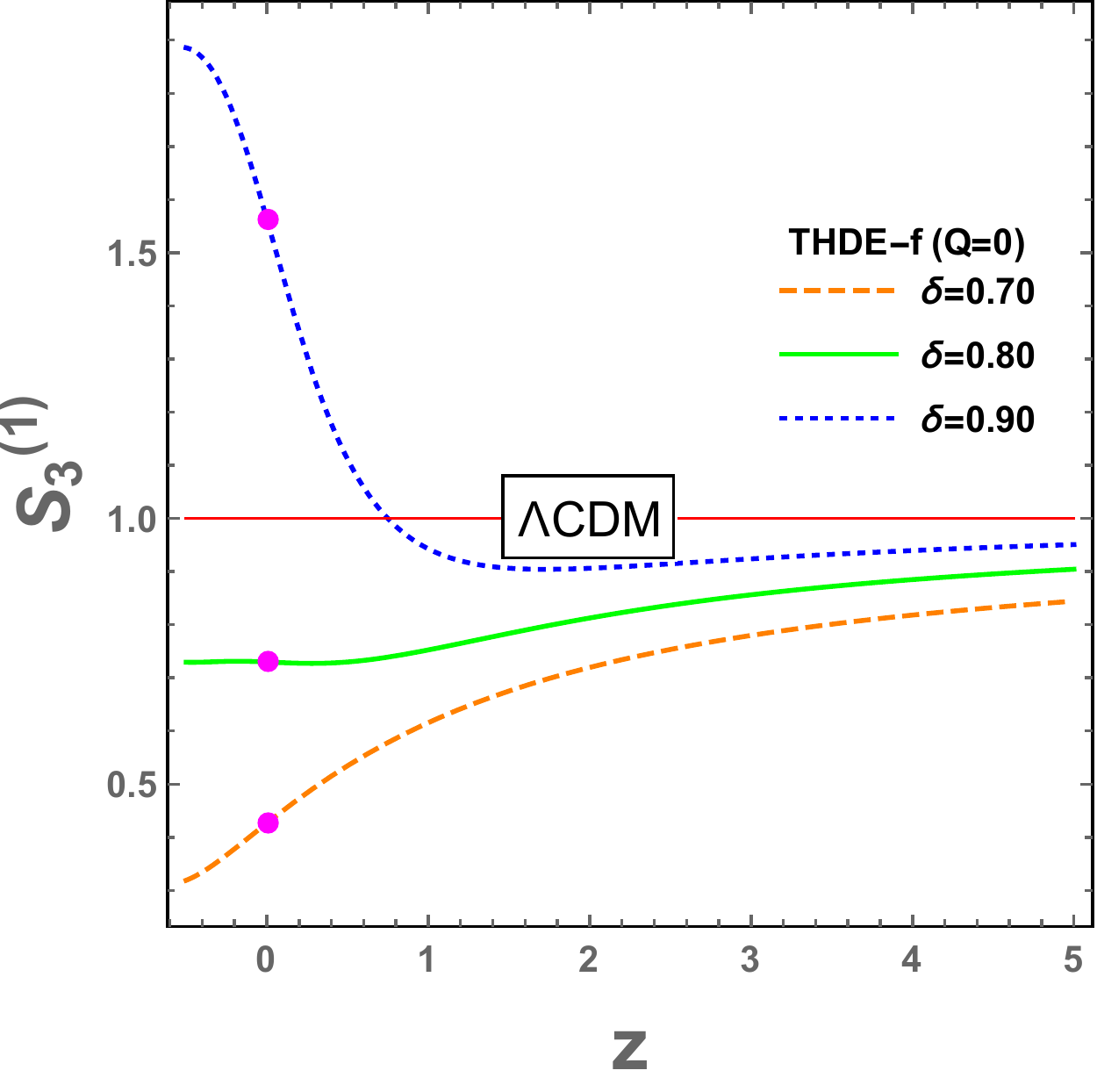}}
\subfigure[]{\label{fig:s3nonc}\includegraphics[width=5cm,height=5cm]{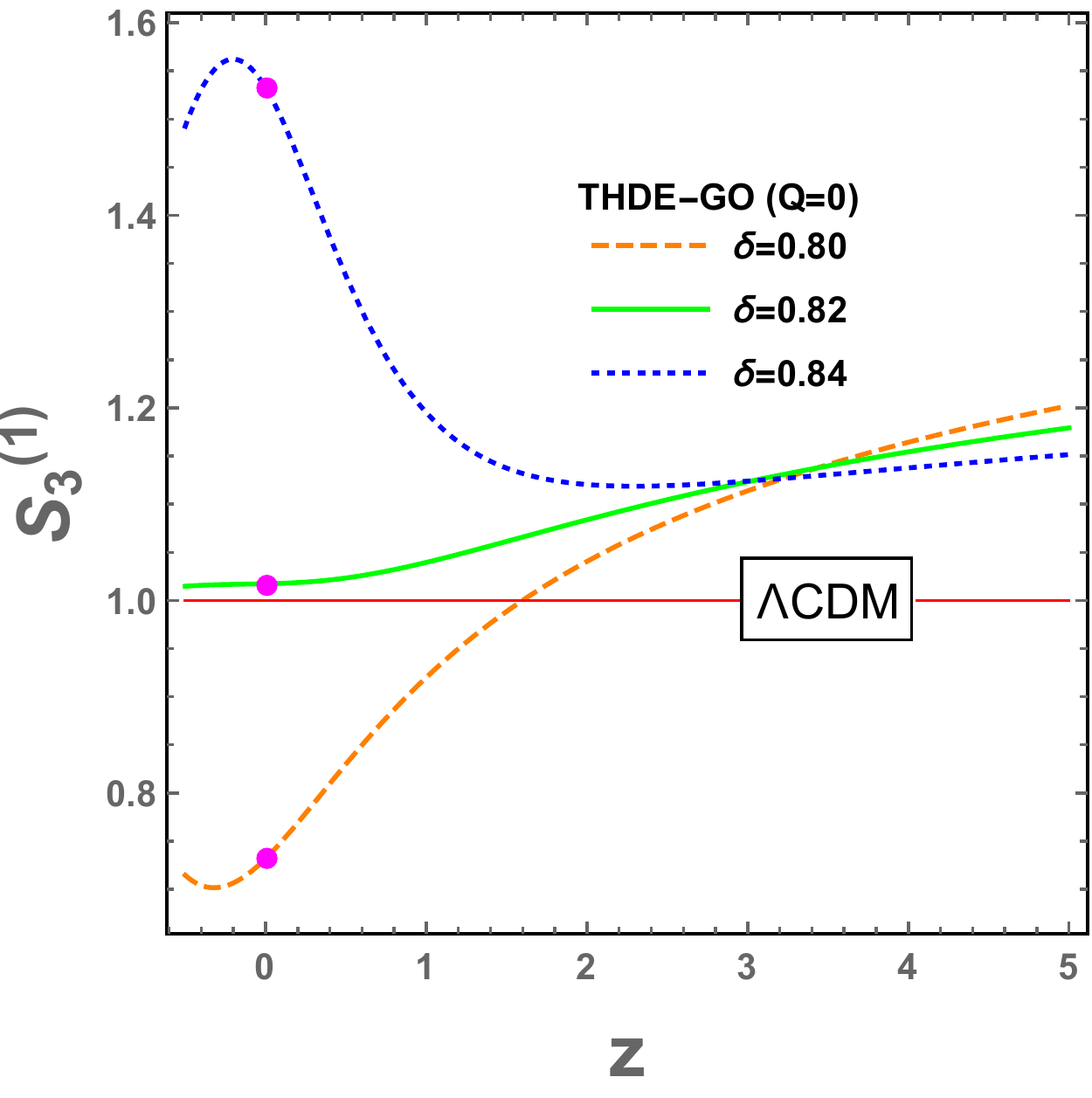}}
\subfigure[]{\label{fig:s3qa}\includegraphics[width=5cm,height=5cm]{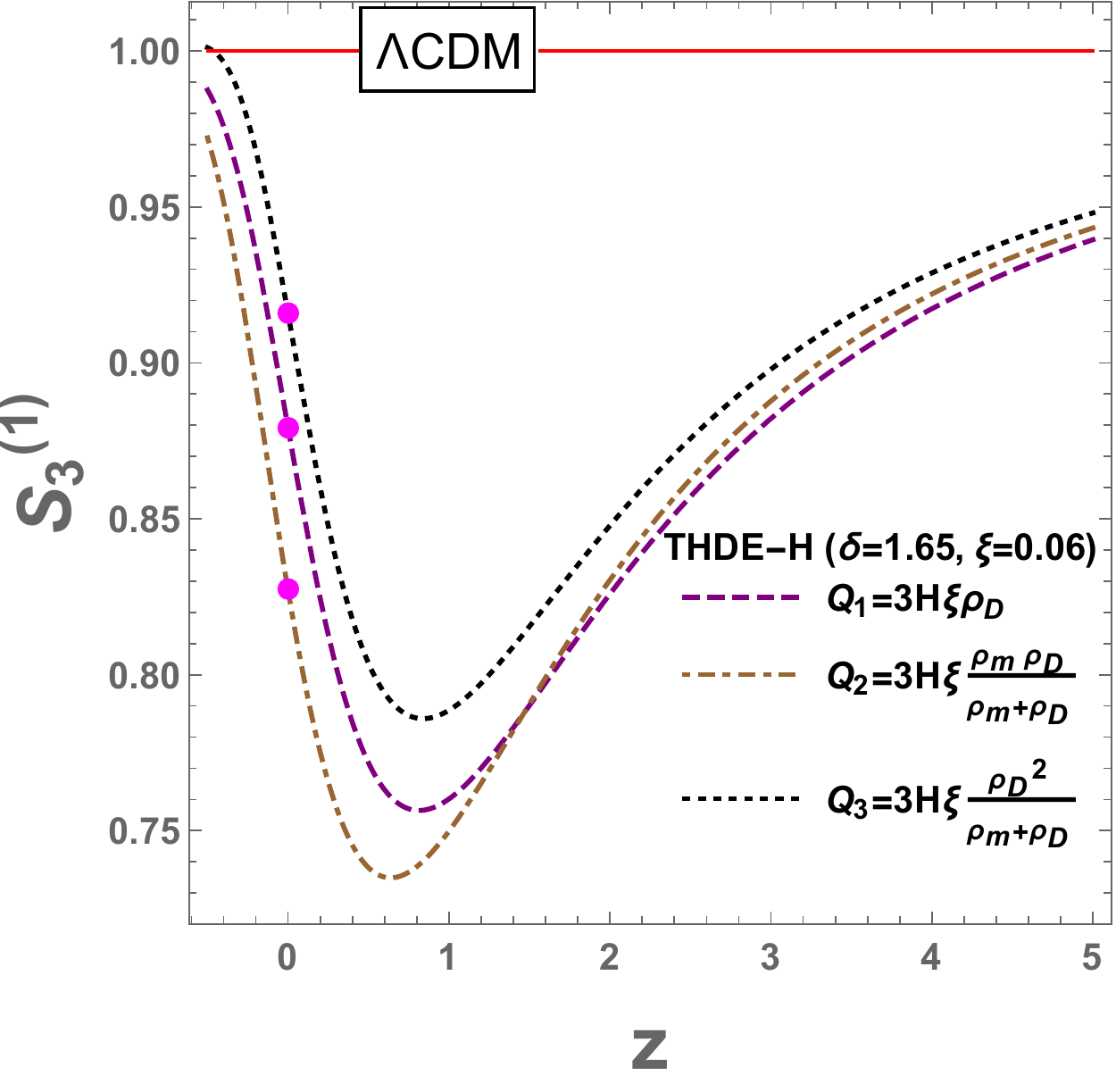}}
\subfigure[]{\label{fig:s3qb}\includegraphics[width=5cm,height=5cm]{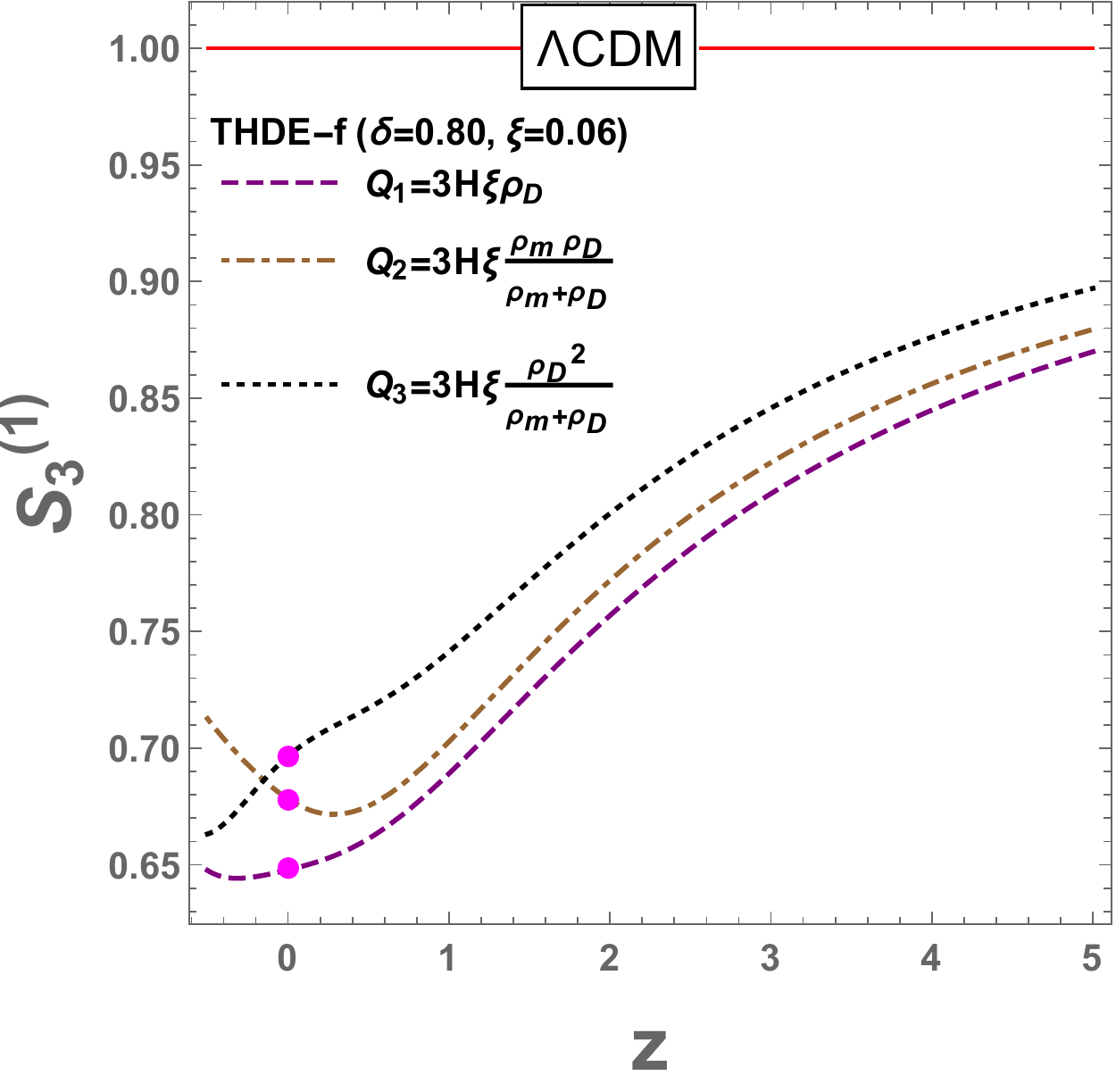}}
\subfigure[]{\label{fig:s3qc}\includegraphics[width=5cm,height=5cm]{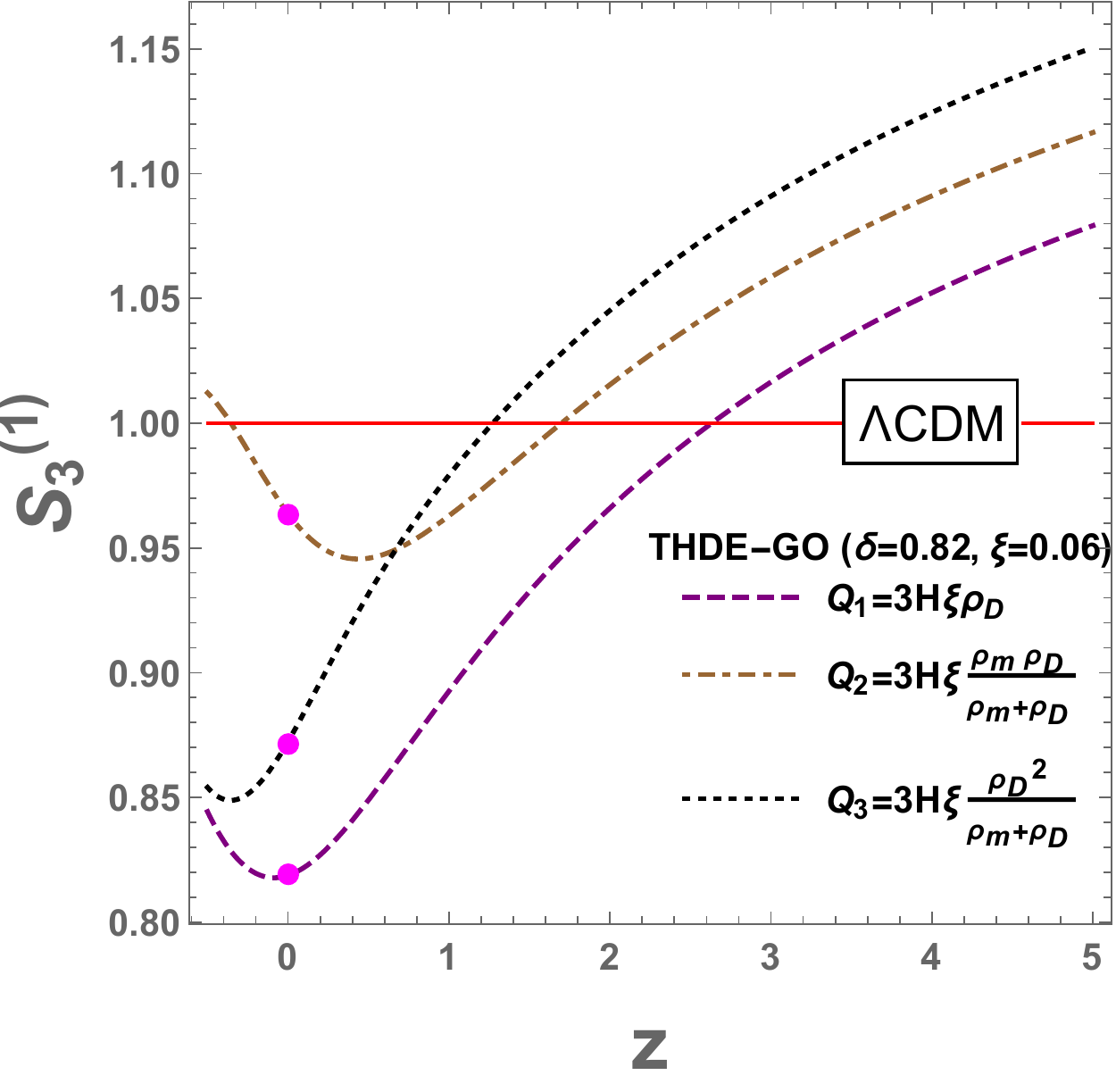}}
\caption{\small{The evolutionary trajectories of $S_{3}^{(1)}$ for THDE models including $Q=0$, where $B=0.8, \alpha=0.9, \beta=0.5.$}}
\label{fig:s3} 
\end{figure}

The evolutions of $S_{3}^{(1)}$ verses to redshift $z$ for THDE models including $Q=0$ are plotted in FIG.~\ref{fig:s3}.
When $Q=0$, from FIGs.~\ref{fig:s3nona}-\ref{fig:s3nonc}, we can find that the influences of model parameter $\delta$ are the same with the ones in $Om$ diagnostic.
When considering the interaction $Q$ in FIGs.~\ref{fig:s3qa}-\ref{fig:s3qc}, the shapes of $S_{3}^{(1)}$ curves have been changed for THDE-H model, the shapes and trends are both changed for THDE-f and THDE-GO models.
Moreover, it can be seen that the statefinder hierarchy $S_{3}^{(1)}$ can distinguish the corresponding evolutionary trajectories to the various values of parameter $\delta$ and the different forms of $Q$ from each other.
But it has to be said that in the future, the curves of different values of $\delta$ for THDE-H model with $Q=0$ are close to each other, and the curve of $Q_{3}$ is close to the horizontal line of $\Lambda$CDM model.

\begin{figure}
\centering
\subfigure[]{\label{fig:s4nona}\includegraphics[width=5cm,height=5cm]{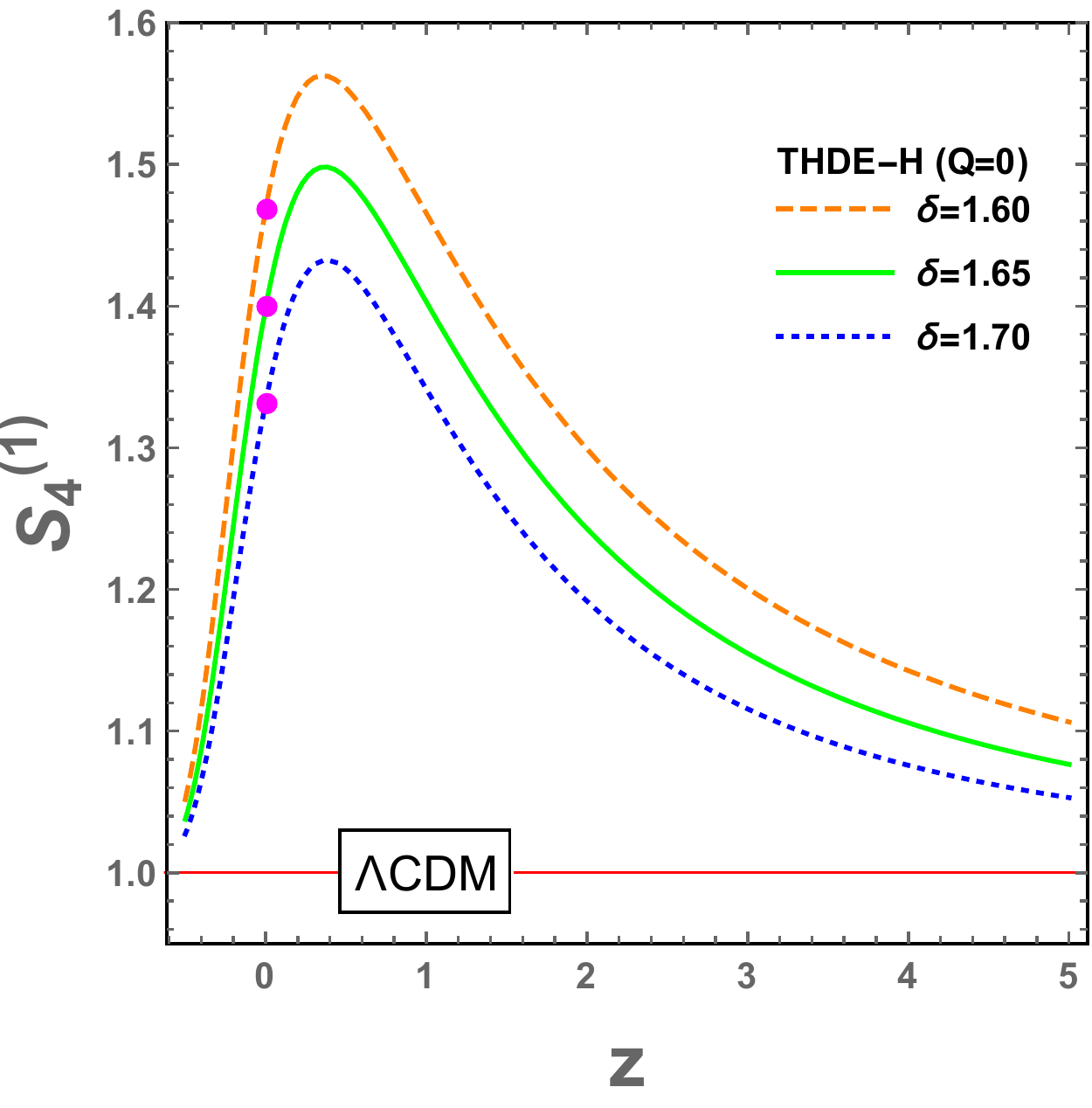}}
\subfigure[]{\label{fig:s4nonb}\includegraphics[width=5cm,height=5cm]{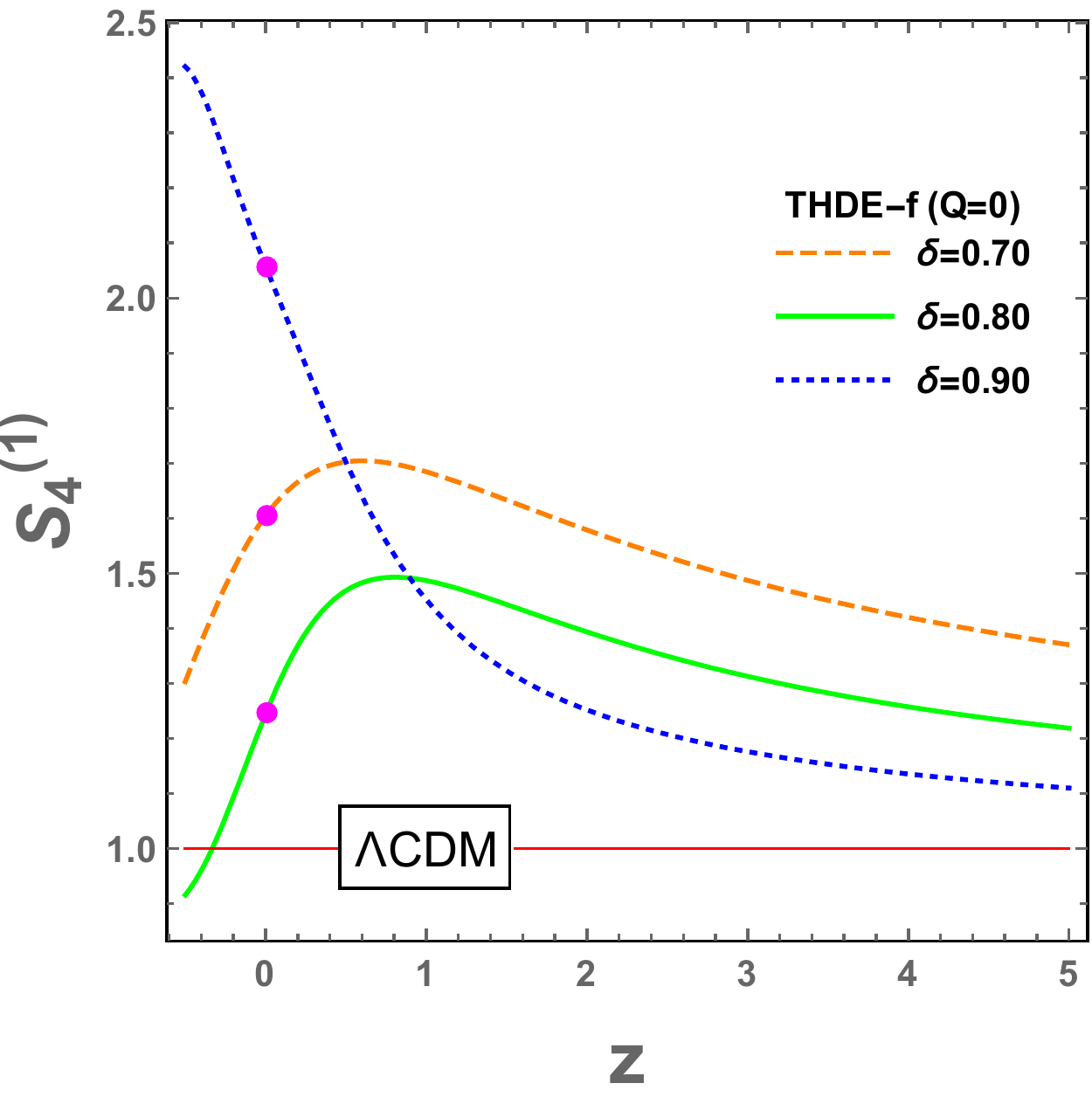}}
\subfigure[]{\label{fig:s4nonc}\includegraphics[width=5cm,height=5cm]{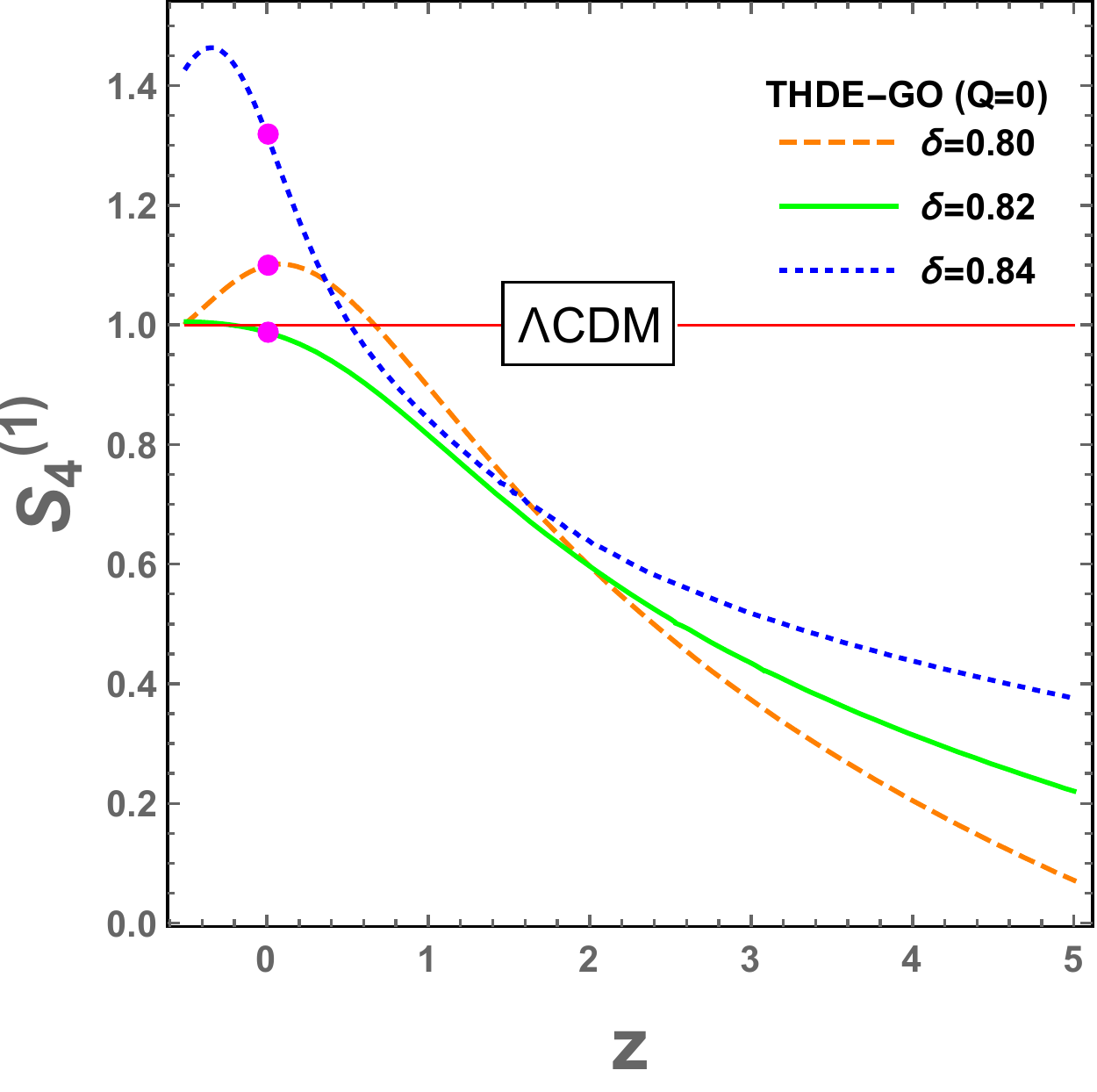}}
\subfigure[]{\label{fig:s4qa}\includegraphics[width=5cm,height=5cm]{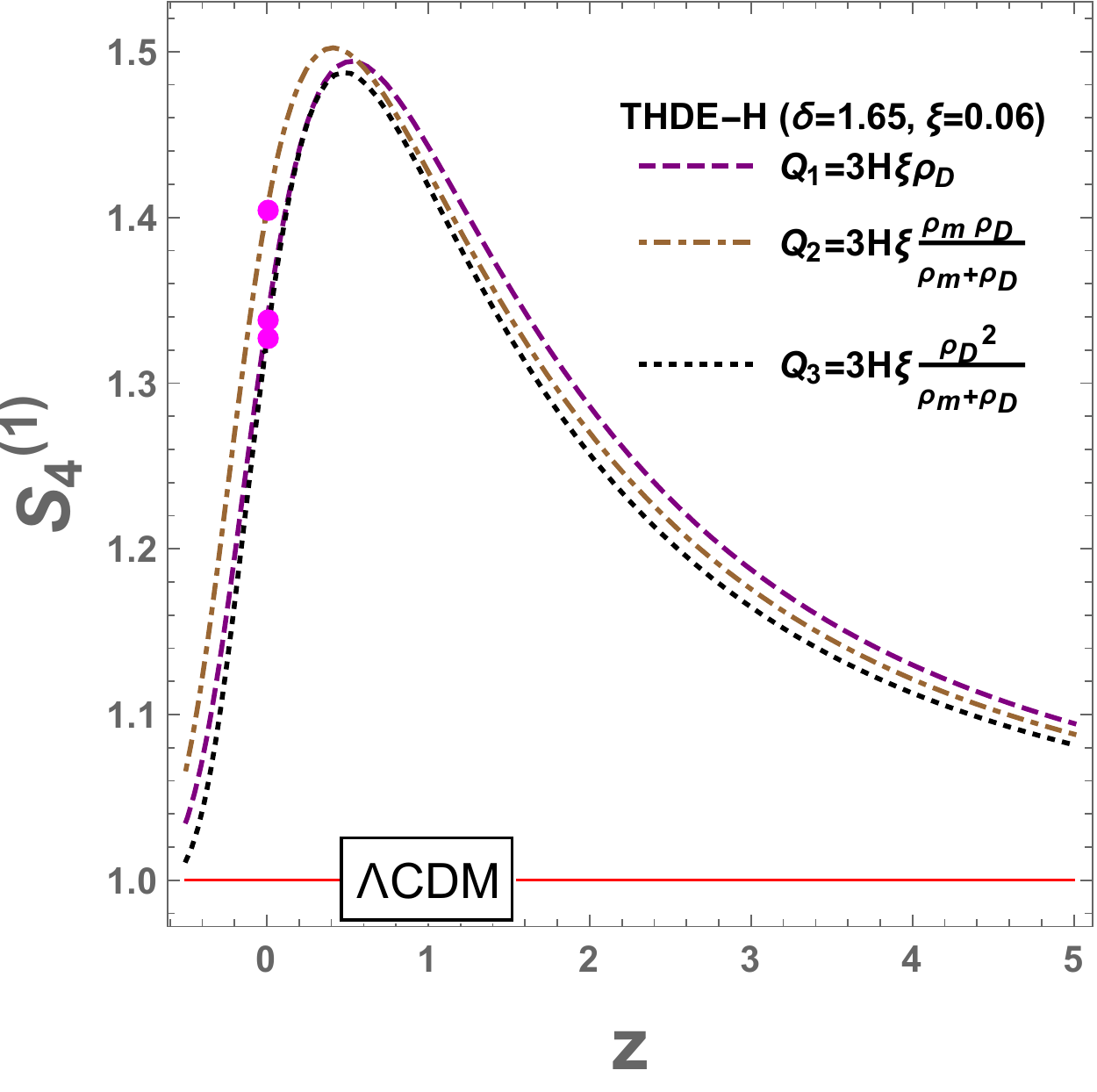}}
\subfigure[]{\label{fig:s4qb}\includegraphics[width=5cm,height=5cm]{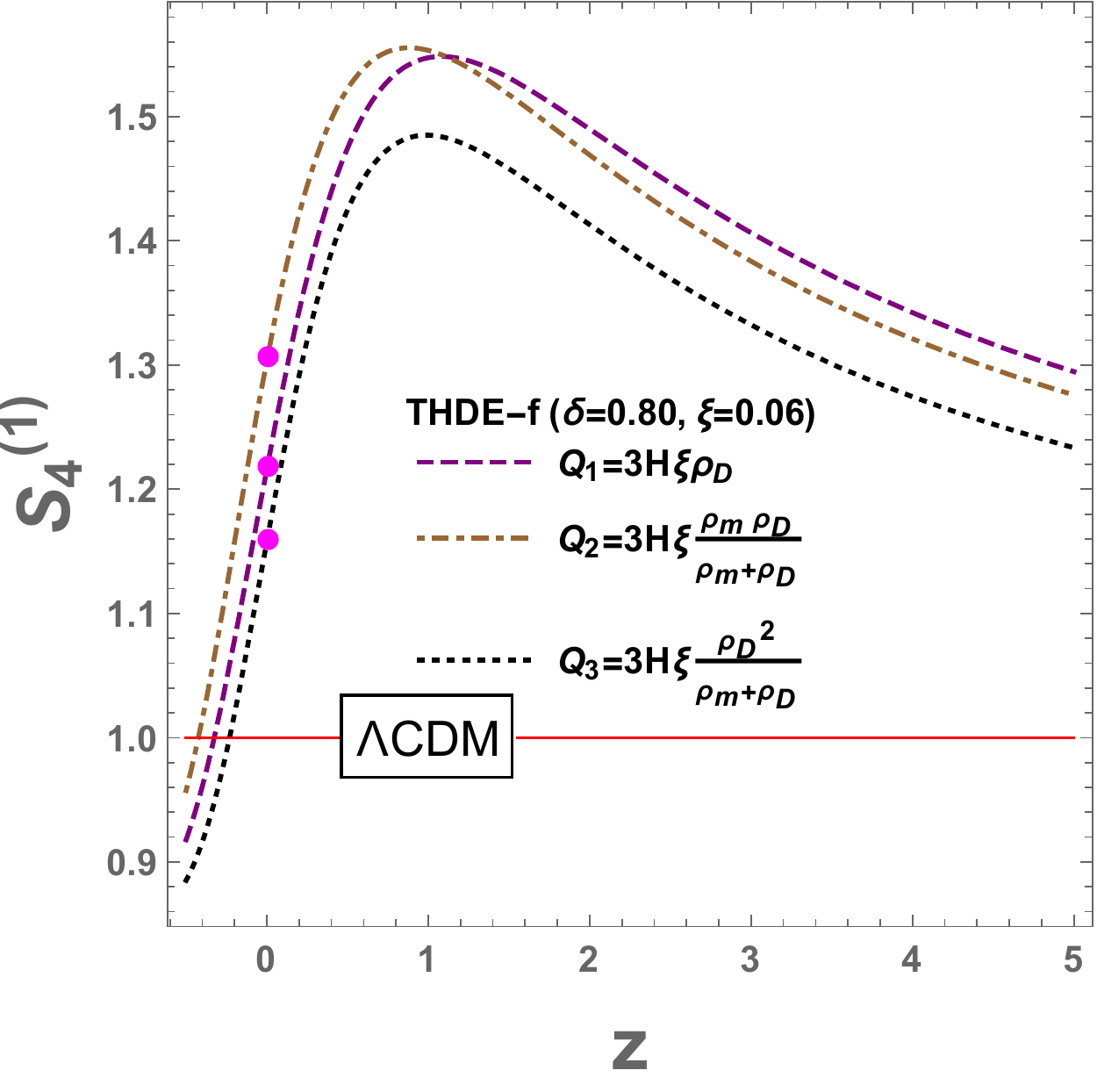}}
\subfigure[]{\label{fig:s4qc}\includegraphics[width=5cm,height=5cm]{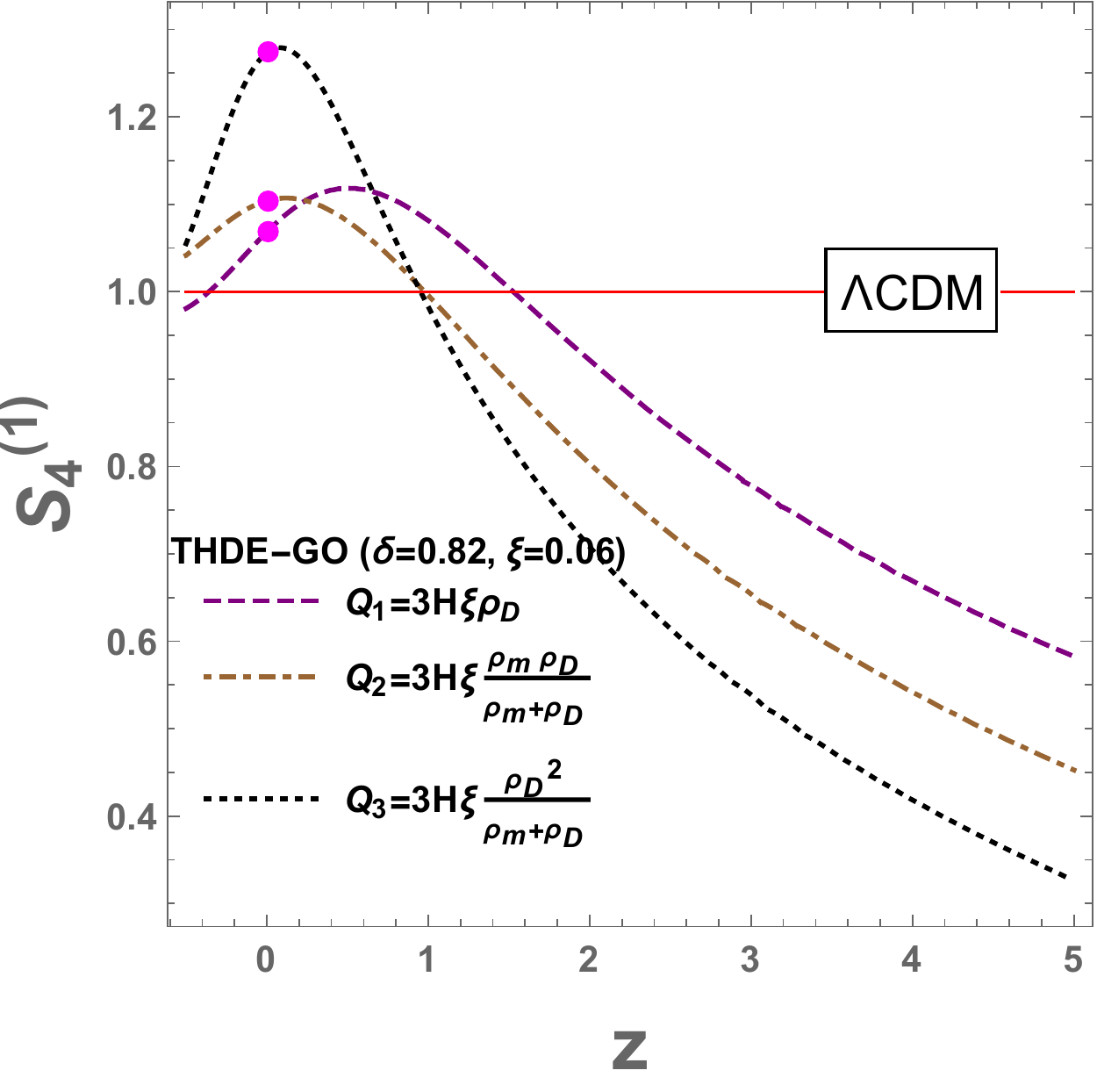}}
\caption{\small{The evolutionary trajectories of $S_{4}^{(1)}$ for THDE models including $Q=0$, where $B=0.8, \alpha=0.9, \beta=0.5.$}}
\label{fig:s4} 
\end{figure}

Similar to $S_{3}^{(1)}$, the evolutions of $S_{4}^{(1)}$ verses to redshift $z$ for THDE models including $Q=0$ are plotted in FIG.~\ref{fig:s4}.
We can also find that the model parameter has the same influence on the model evolutions (see FIGs.~\ref{fig:s4nona}-\ref{fig:s4nonc}).
Besides, the existences of $Q$ change the trends of $S_{4}^{(1)}$ curves only for THDE-GO model in FIG.~\ref{fig:s4qc} and the interactions could also break the degeneracy of THDE-GO model with $\Lambda$CDM model in FIG.~\ref{fig:s4nonc}.
Moreover, in the aspect of diagnosing THDE models by $S_{4}^{(1)}$, the curves of different values of $\delta$ for THDE-H model are close to each other in the future, and the curves of $Q_{1}$ and $Q_{3}$ are nearly overlapped in the low-redshift and future regions.

\begin{table}[h]
\vspace{-0.5cm}
\centering
\tbl{\small{The present values of the statefinders $S_{3to}^{(1)}$, $S_{4to}^{(1)}$ and the fractional growth parameters $\epsilon_{to}$, and the differences of them, $\triangle S_{3to}^{(1)}$, $\triangle S_{4to}^{(1)}$ and $\triangle \epsilon_{to}$, for THDE models where $B=0.8, \alpha=0.9, \beta=0.5$ and $\xi=0.06$.}}
{\resizebox{16cm}{1.8cm}{
\begin{tabular}{|c|ccc|ccc|ccc|ccc|ccc|ccc|}
\hline
$Parameters$     & \multicolumn{6}{c|} {$THDE-H$}   & \multicolumn{6}{c|} {$THDE-f$}  & \multicolumn{6}{c|} {$THDE-GO$}\\ \cline{2-19}
 $Q$      & \multicolumn{3}{c|} {$Q=0$}        & $Q_{1}$ &  $Q_{2}$  & $Q_{3}$
          & \multicolumn{3}{c|} {$Q=0$}       & $Q_{1}$ &  $Q_{2}$  & $Q_{3}$
          & \multicolumn{3}{c|} {$Q=0$}       & $Q_{1}$ &  $Q_{2}$  & $Q_{3}$  \\ \cline{2-19}
 $\delta$        & $1.60$ &  $1.65$  & $1.70$  & \multicolumn{3}{c|} {$\delta=1.65$}
                 & $0.70$ &  $0.80$  & $0.90$  & \multicolumn{3}{c|}{$\delta=0.80$}
                 & $0.80$ &  $0.82$  & $0.84$  &\multicolumn{3}{c|}{$\delta=0.82$}\\ \hline
$S_{3to}^{(1)}$   & $0.838$           &$0.868$     & $0.894$  & $0.879$           &$0.827$       & $0.916$
                  & $0.426$           &$0.730$       & $1.564$& $0.648$           &$0.678$       & $0.697$
                  & $0.732$           &$1.017$       & $1.533$  & $0.819$        &$0.964$       & $0.872$  \\
$S_{4to}^{(1)}$   & $1.469$           &$1.399$       & $1.332$ & $1.338$           &$1.405$       & $1.327$
                  & $1.605$           &$1.246$       & $2.057$& $1.218$           &$1.306$       & $1.160$
                   & $1.099$           &$0.988$       & $1.318$ & $1.068$        &$1.104$       & $1.273$\\
$\epsilon_{to}$   & $0.989$           &$0.991$       & $0.993$
                  & $0.974$           &$0.986$       & $0.998$
                   & $0.978$           &$0.988$       & $0.997$
                   & $0.926$           &$0.966$       & $0.952$
                  & $0.914$           &$0.959$       & $0.943$
                   & $0.913$           &$0.959$       & $0.945$\\
$\triangle S_{3to}^{(1)}$         & \multicolumn{3}{c|}{$0.056$}  & \multicolumn{3}{c|}{$0.089$} & \multicolumn{3}{c|}{$1.138$} & \multicolumn{3}{c|}{$0.049$} & \multicolumn{3}{c|}{$0.801$}    & \multicolumn{3}{c|}{$0.145$}\\
$\triangle S_{4to}^{(1)}$         & \multicolumn{3}{c|}{$0.137$} & \multicolumn{3}{c|}{$0.078$} & \multicolumn{3}{c|}{$0.811$}  & \multicolumn{3}{c|}{$0.146$}& \multicolumn{3}{c|}{$0.330$}    & \multicolumn{3}{c|}{$0.205$} \\
$\triangle \epsilon_{to}$         & \multicolumn{3}{c|}{$0.004$}  & \multicolumn{3}{c|}{$0.024$} & \multicolumn{3}{c|}{$0.019$}  & \multicolumn{3}{c|}{$0.040$}  & \multicolumn{3}{c|}{$0.045$} & \multicolumn{3}{c|}{$0.046$}\\
\hline
\end{tabular}\label{tab:s}}}
\end{table}

We give the present values of $S_{3to}^{(1)}$, $S_{4to}^{(1)}$ and $\triangle S_{3to}^{(1)}$, $\triangle S_{4to}^{(1)}$ for interacting THDE models including $Q=0$ in Table~\ref{tab:s}, where $\triangle S_{3to}^{(1)}=S_{3to}^{(1)}(max)-S_{3to}^{(1)}(min)$, $\triangle S_{4to}^{(1)}=S_{4to}^{(1)}(max)-S_{4to}^{(1)}(min)$ and the subscripts $to$ means $today$.
The current values also play an important role in diagnosing DE models.
As we can see, for THDE-H model, $\triangle S_{3to}^{(1)}<\triangle S_{4to}^{(1)}$ when $Q=0$ and $\triangle S_{3to}^{(1)}>\triangle S_{4to}^{(1)}$ for different forms of $Q$, which also indicate that $S_{4}^{(1)}$ can give larger difference among the curves corresponding to the different values of $\delta$ than the one given by $S_{3}^{(1)}$, on the contrary, $S_{3}^{(1)}$ will make us easier to distinguish the different forms of $Q$ compared with $S_{4}^{(1)}$.
This kind of phenomenon, i.e., larger difference among the curves can give better diagnostic result, is also appropriate for diagnosing different values of $\delta$ in THDE-f and THDE-GO models, in which $\triangle S_{3to}^{(1)}>\triangle S_{4to}^{(1)}$ and $S_{3}^{(1)}$ can give better diagnostic results according to these figures.
But we can find that $\triangle S_{3to}^{(1)}<\triangle S_{4to}^{(1)}$ in THDE-f and THDE-GO models when $Q\neq0$, while these figures imply that compared with $S_{4}^{(1)}$ in FIGs.~\ref{fig:s4qb}, \ref{fig:s4qc}, $S_{3}^{(1)}$ can better distinguish different forms of $Q$ (see FIGs.~\ref{fig:s3qb}, \ref{fig:s3qc}).
Thus, the magnitudes of $\triangle S_{3to}^{(1)}$ and $\triangle S_{4to}^{(1)}$ can not determine the effectiveness of $S_{3}^{(1)}$ and $S_{4}^{(1)}$ when $Q\neq0$.

\begin{figure}
\centering
\subfigure[]{\label{fig:s3e0a}\includegraphics[width=5cm,height=5cm]{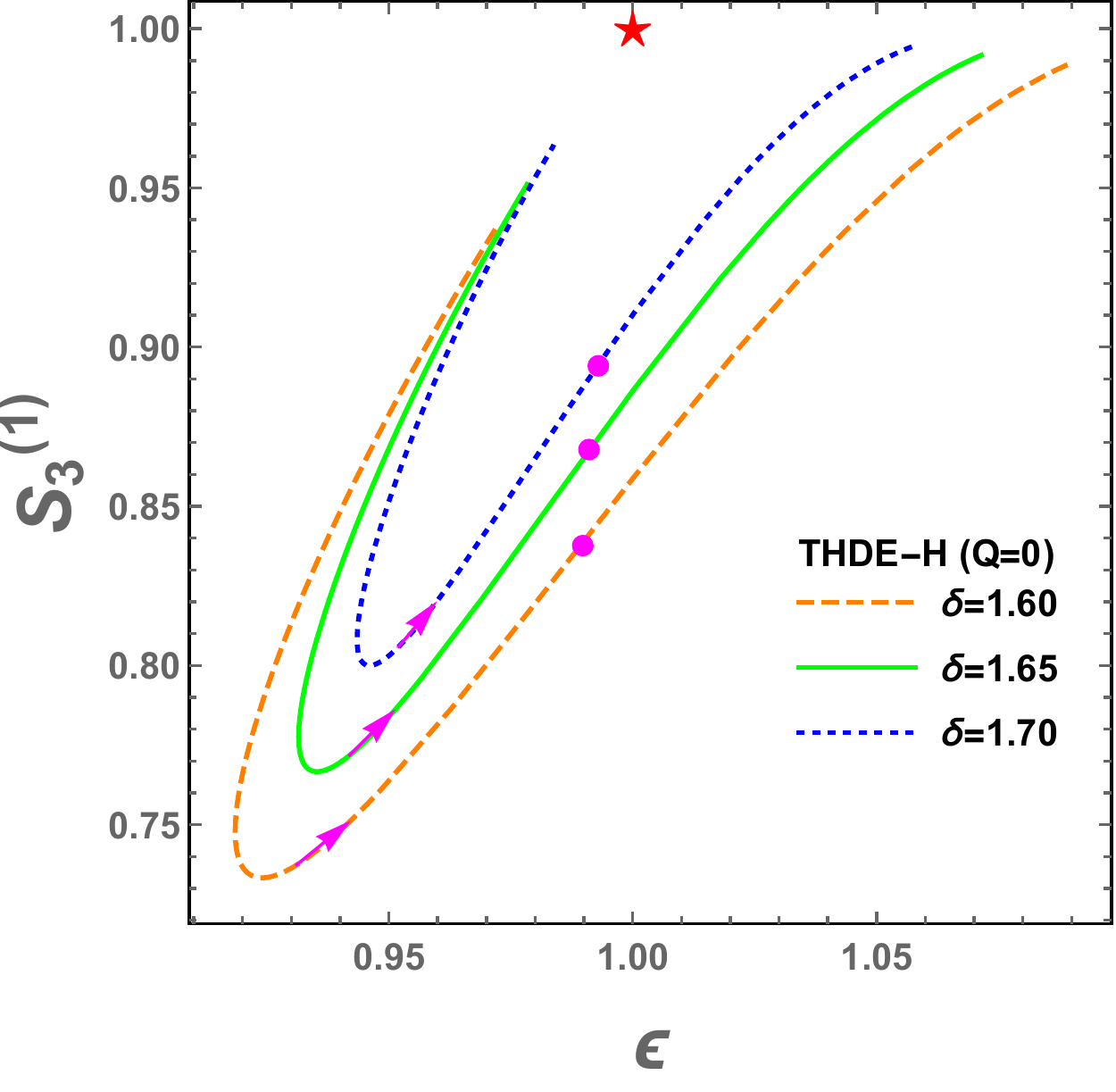}}
\subfigure[]{\label{fig:s3e0b}\includegraphics[width=5cm,height=5cm]{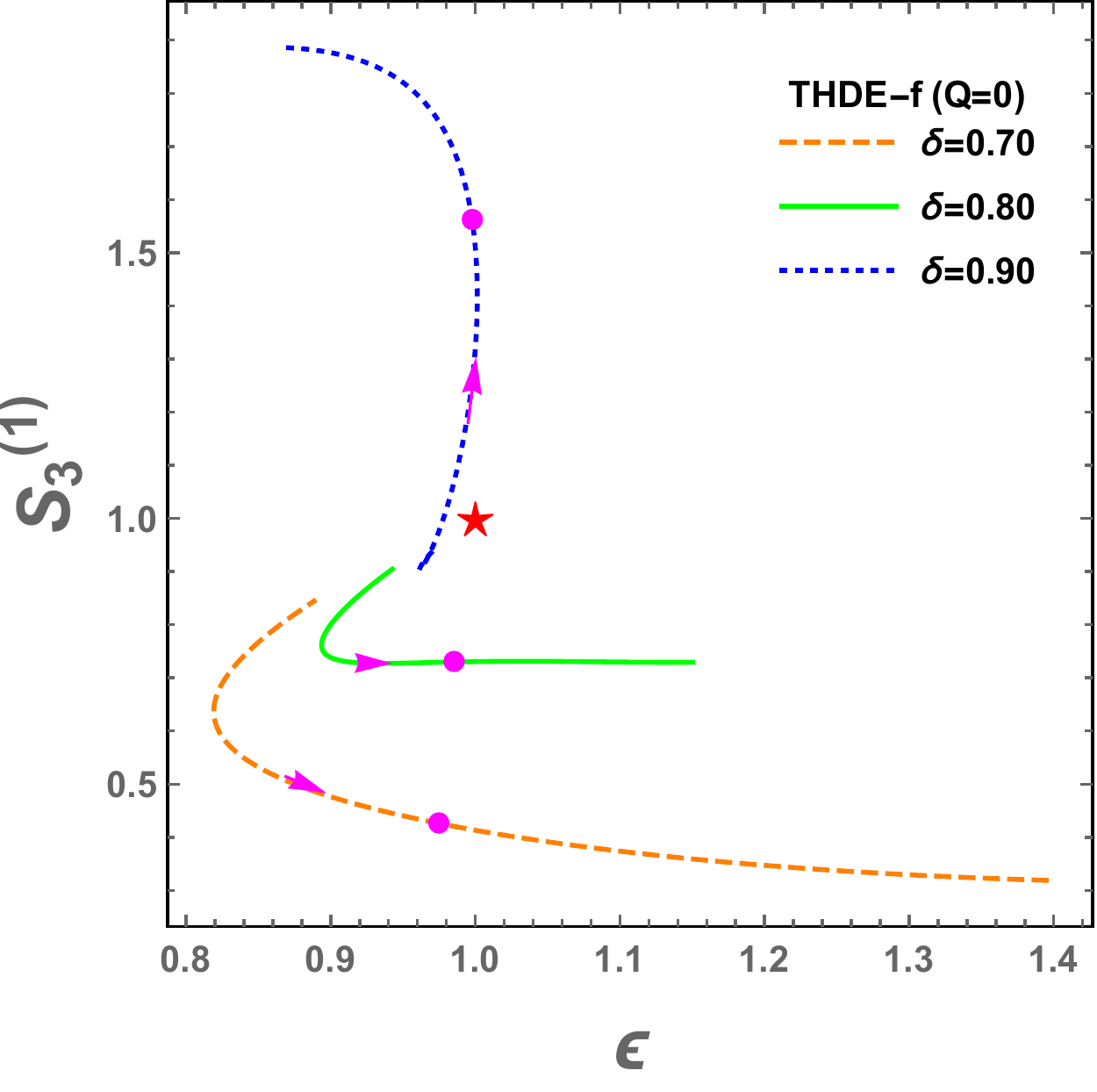}}
\subfigure[]{\label{fig:s3e0c}\includegraphics[width=5cm,height=5cm]{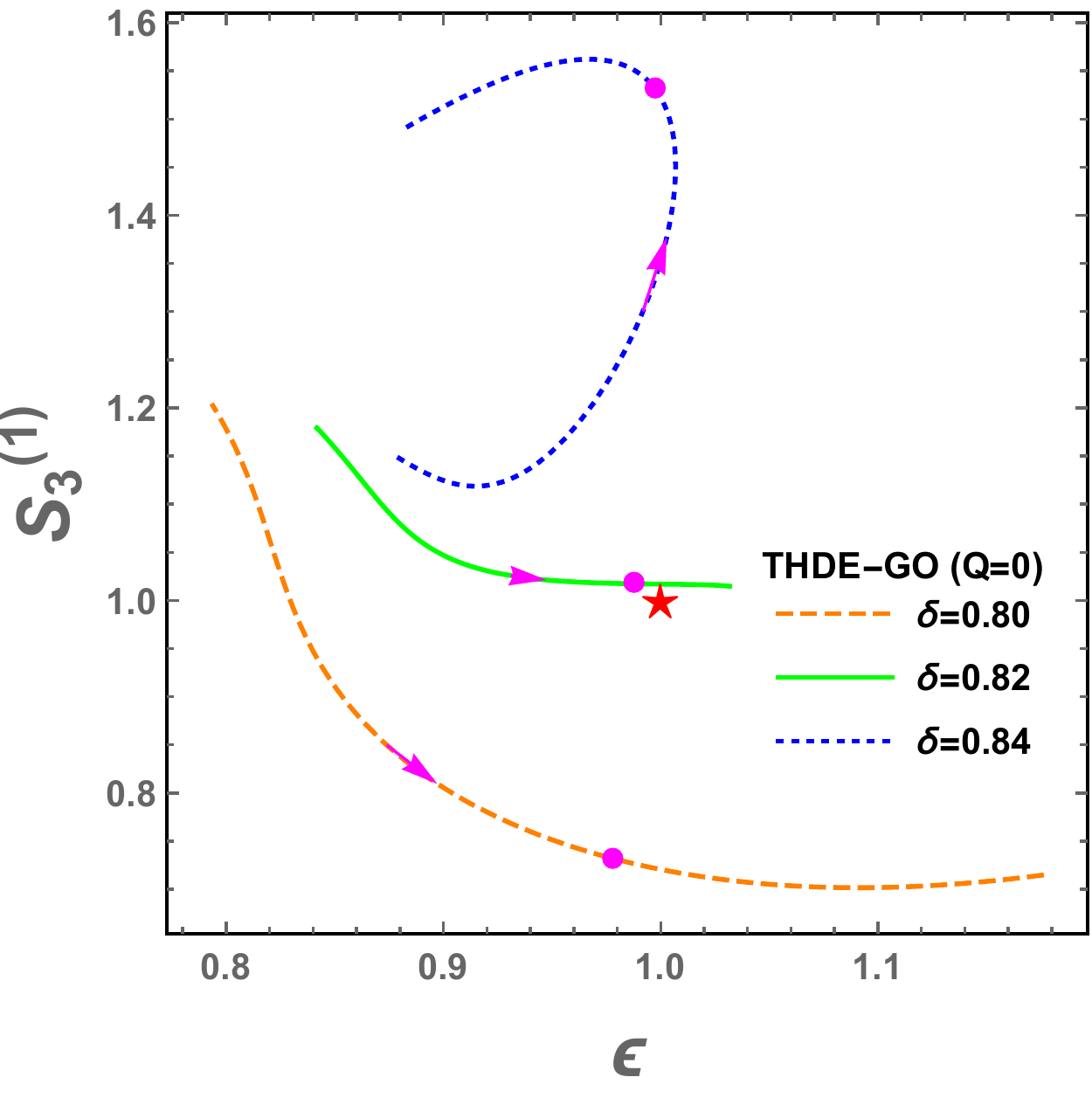}}
\subfigure[]{\label{fig:s3eqa}\includegraphics[width=5cm,height=5cm]{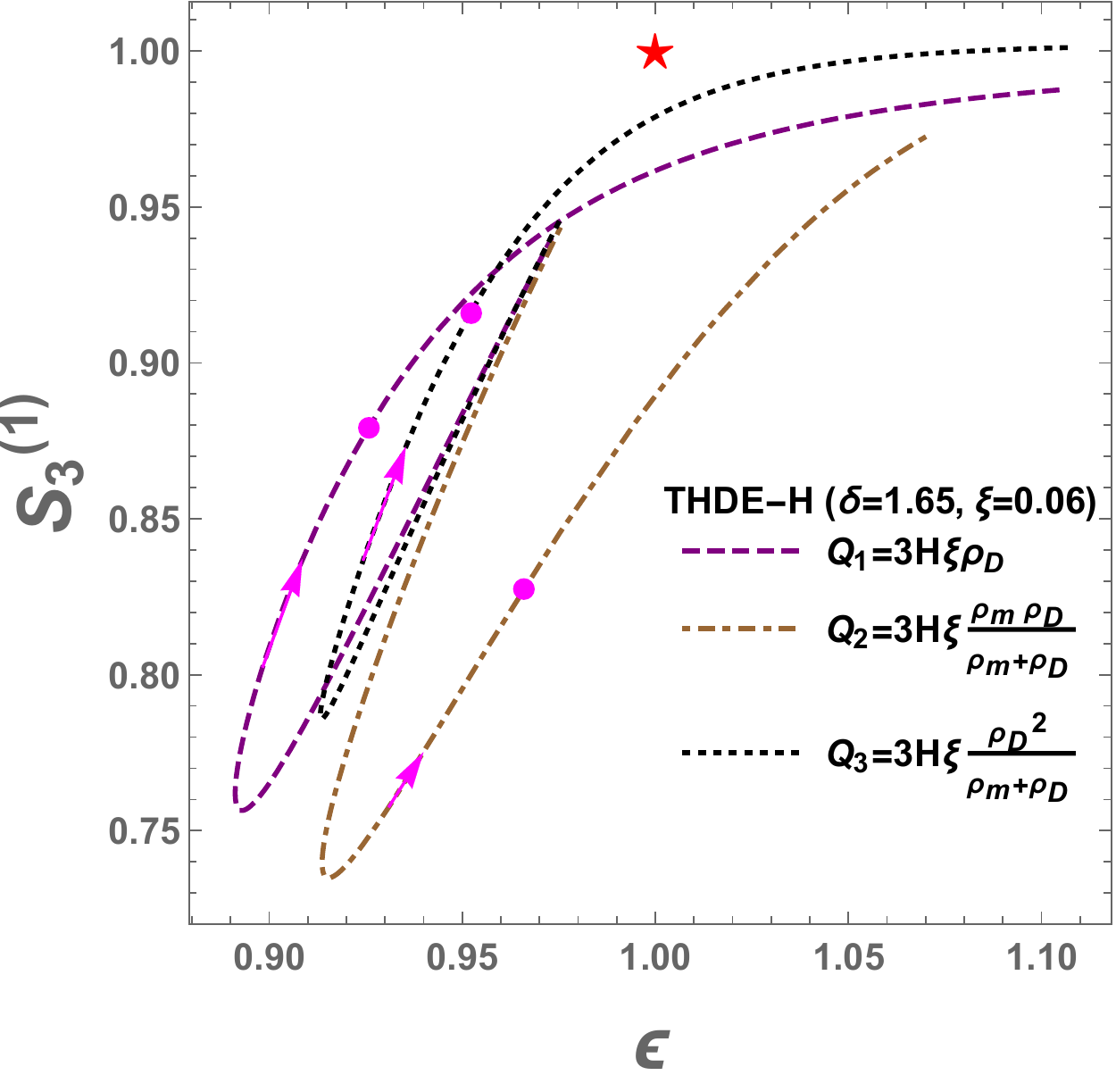}}
\subfigure[]{\label{fig:s3eqb}\includegraphics[width=5cm,height=5cm]{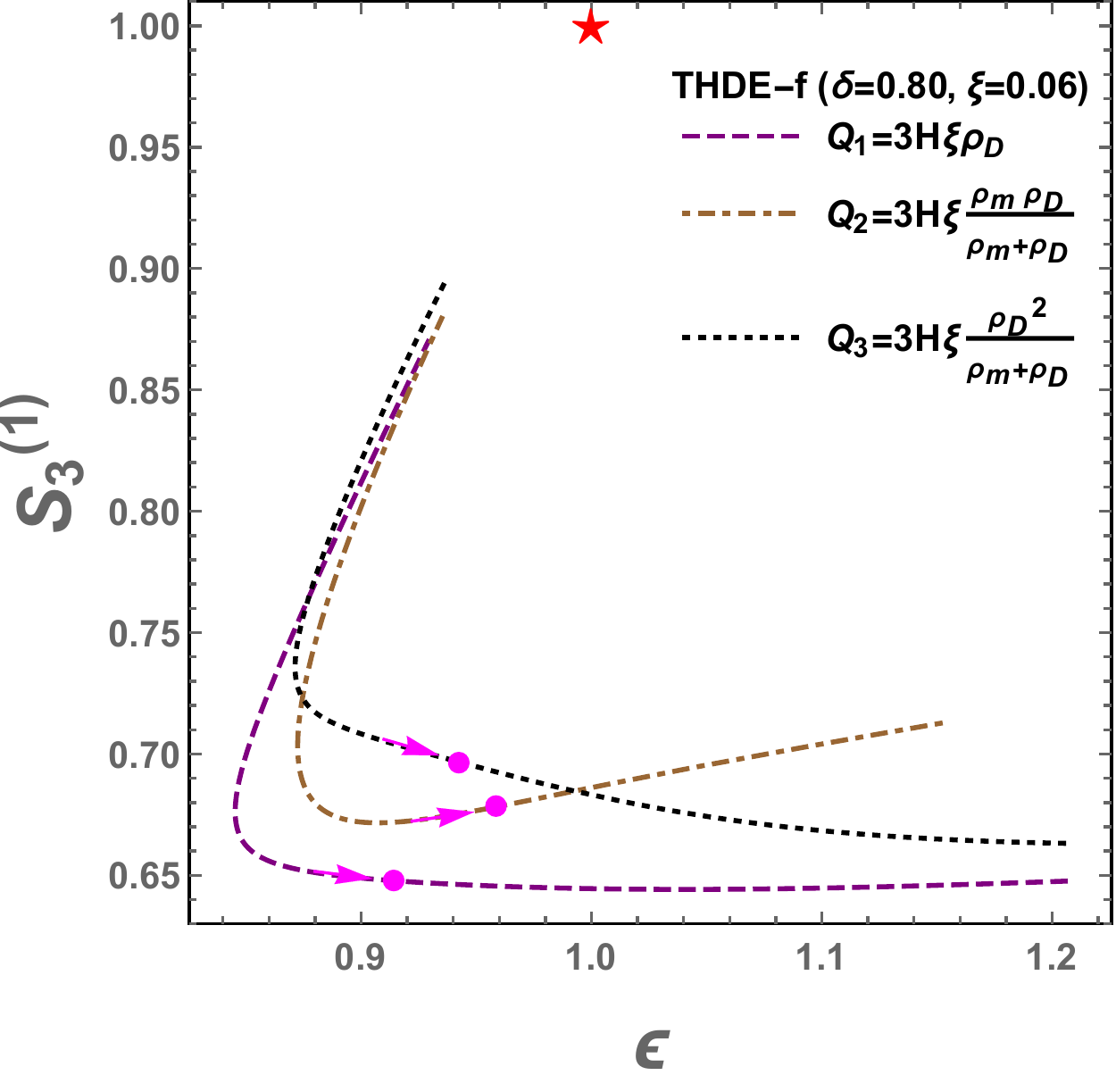}}
\subfigure[]{\label{fig:s3eqc}\includegraphics[width=5cm,height=5cm]{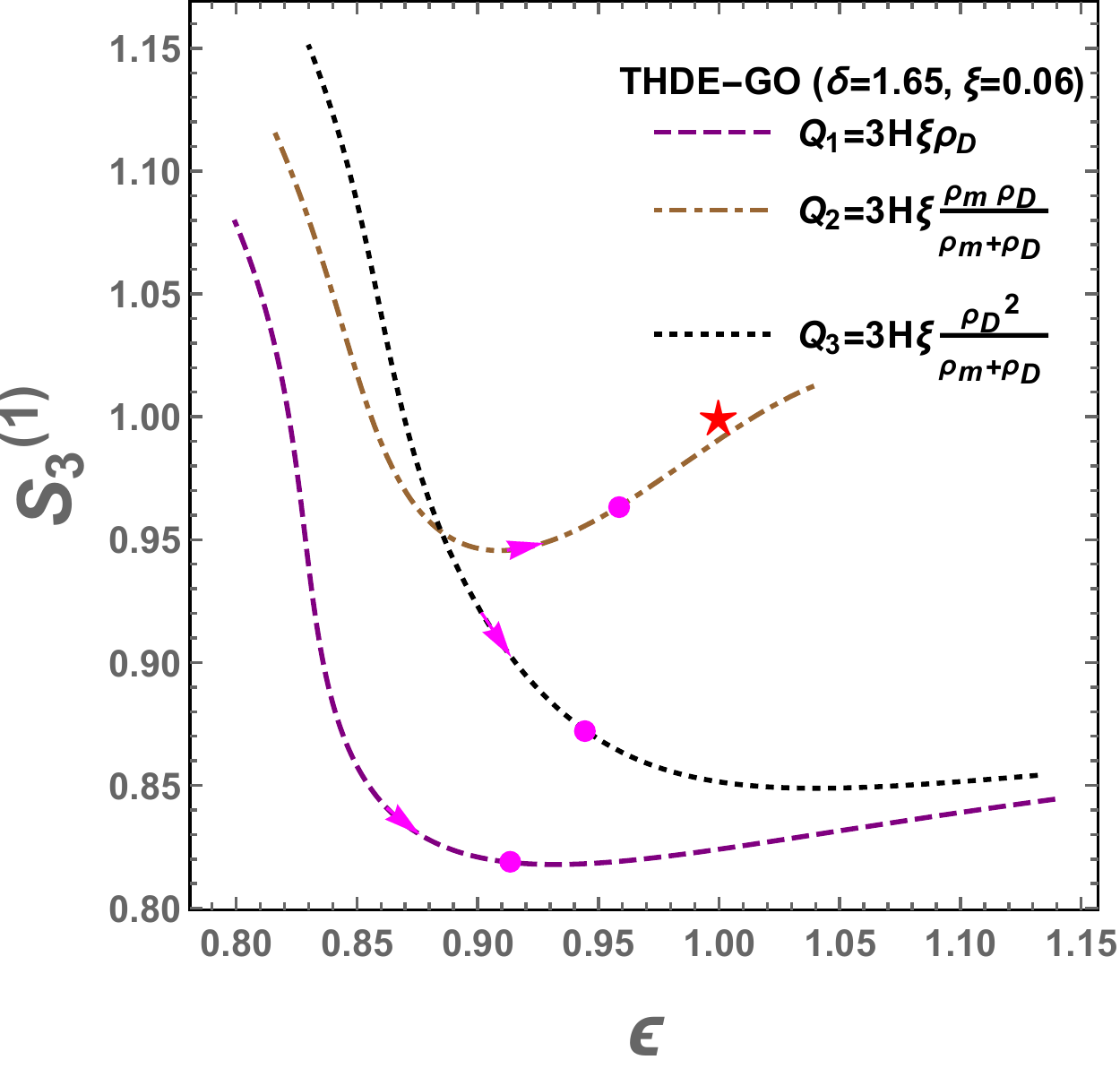}}
\caption{\small{The evolutionary trajectories of CND $\{S_{3}^{(1)},\epsilon\}$ for THDE models including $Q=0$, where $B=0.8, \alpha=0.9, \beta=0.5.$}}
\label{fig:se3} 
\end{figure}

\begin{figure}
\centering
\subfigure[]{\label{fig:s4e0a}\includegraphics[width=5cm,height=5cm]{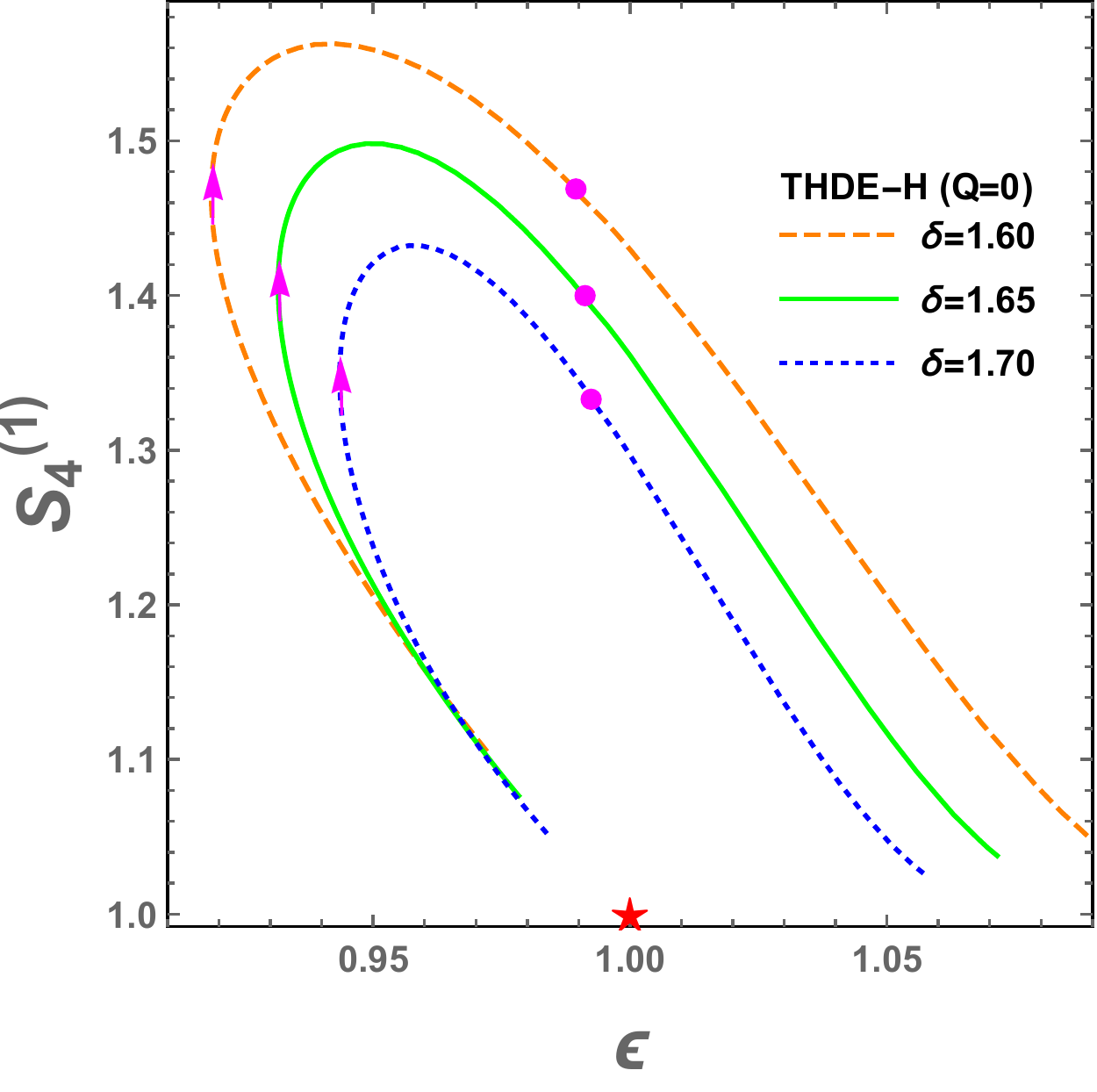}}
\subfigure[]{\label{fig:s4e0b}\includegraphics[width=5cm,height=5cm]{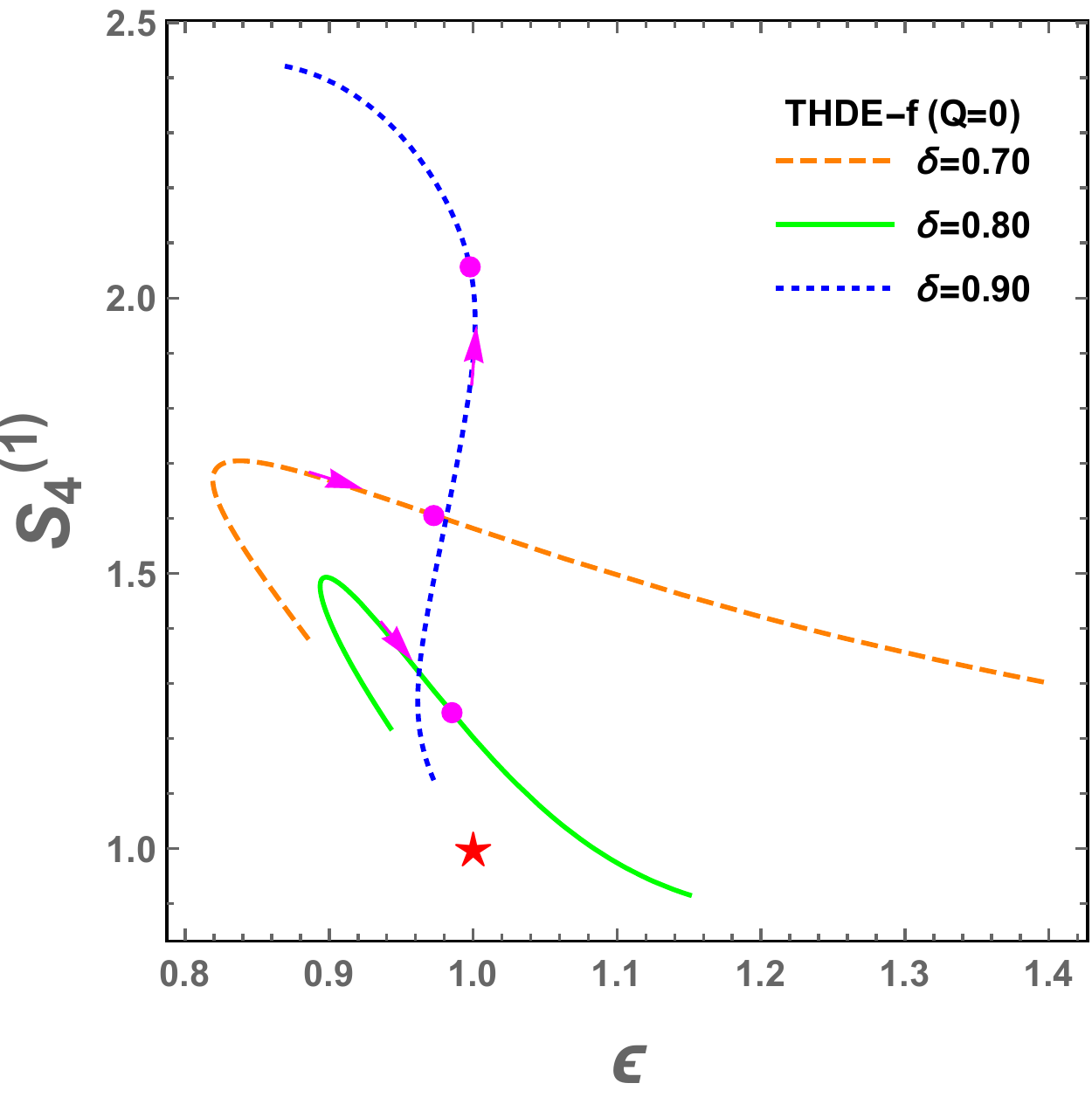}}
\subfigure[]{\label{fig:s4e0c}\includegraphics[width=5cm,height=5cm]{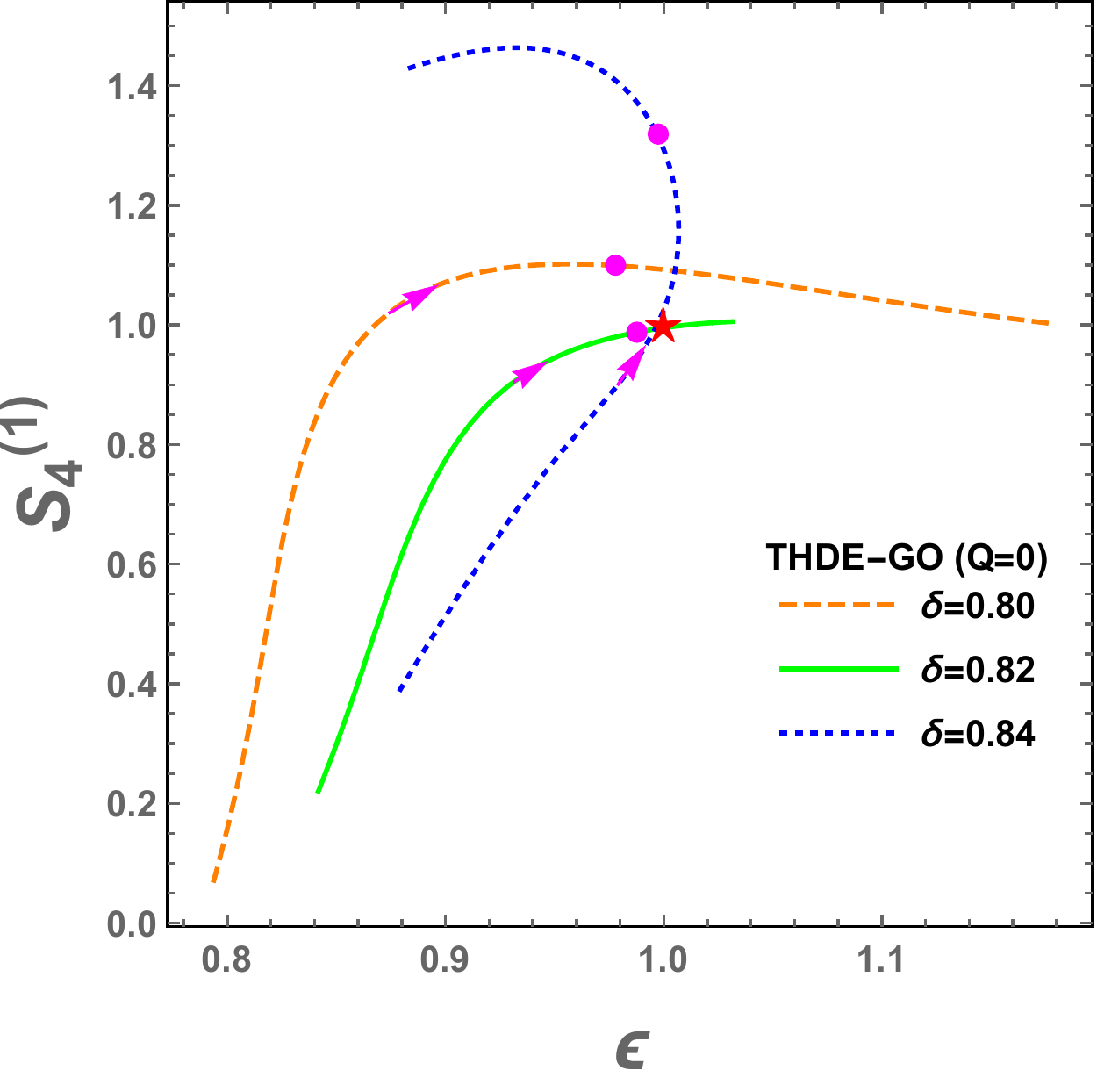}}
\subfigure[]{\label{fig:s4eqa}\includegraphics[width=5cm,height=5cm]{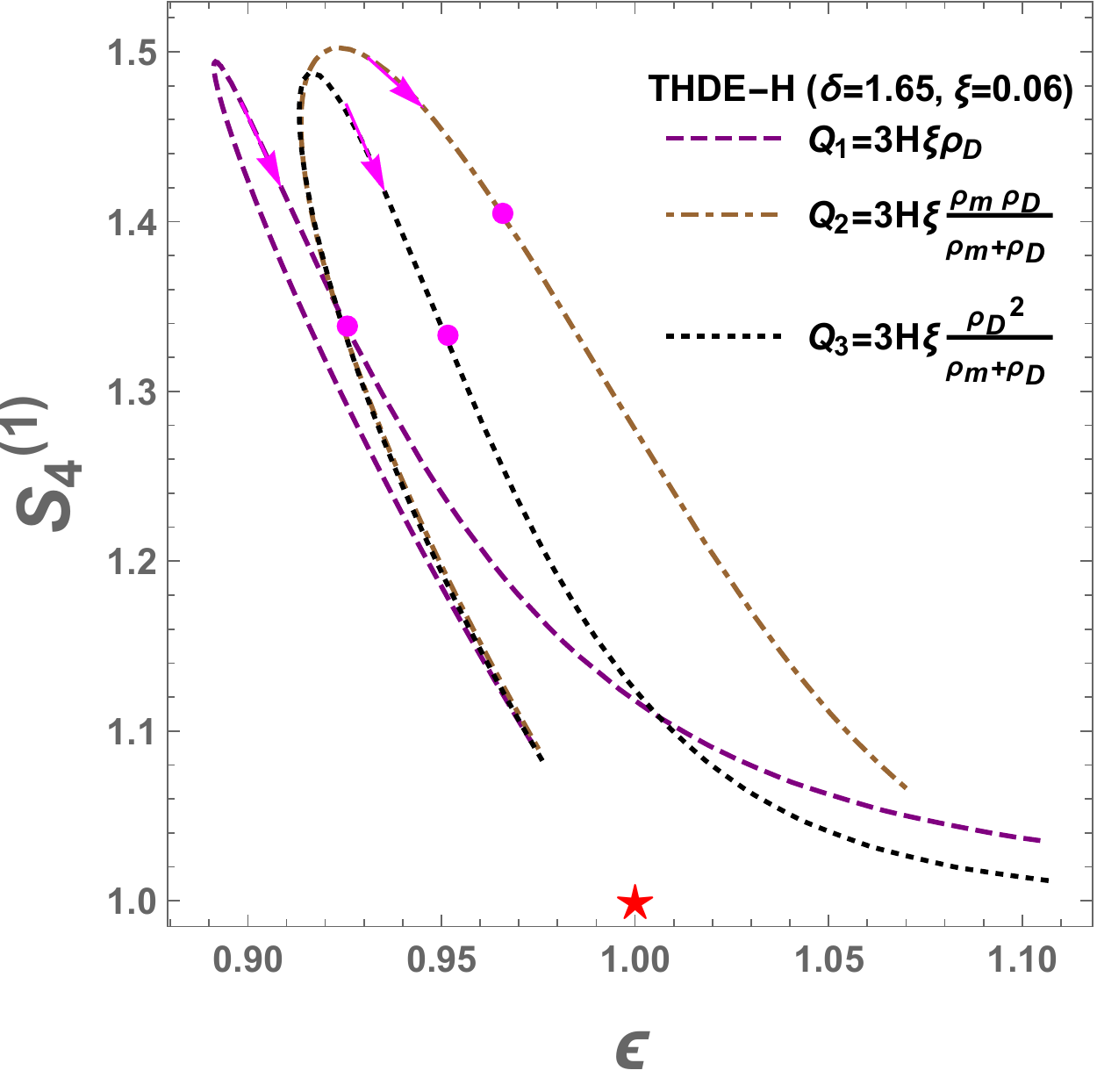}}
\subfigure[]{\label{fig:s4eqb}\includegraphics[width=5cm,height=5cm]{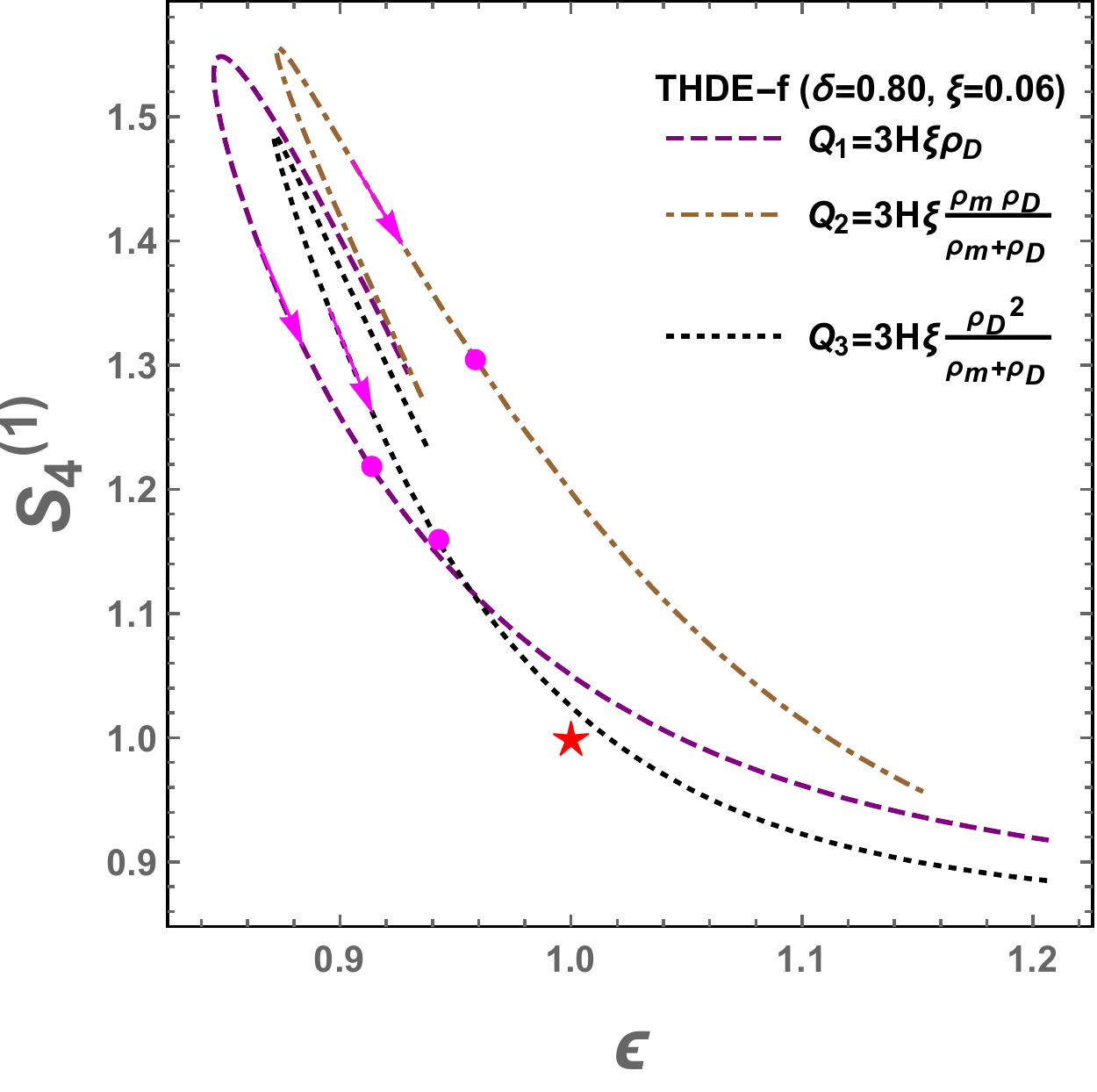}}
\subfigure[]{\label{fig:s4eqc}\includegraphics[width=5cm,height=5cm]{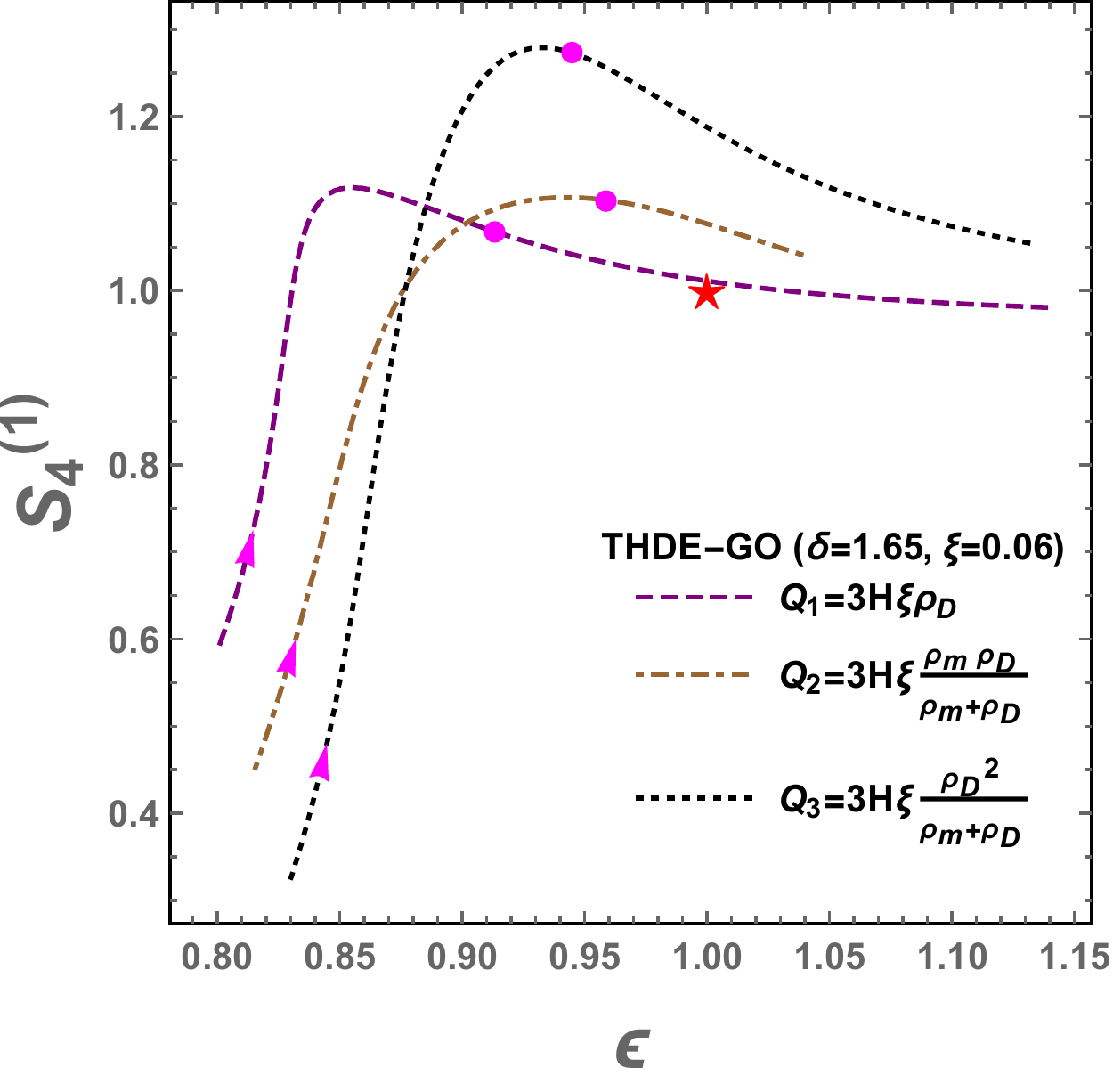}}
\caption{\small{The evolutionary trajectories of CND $\{S_{4}^{(1)},\epsilon\}$ for THDE models including $Q=0$, where $B=0.8, \alpha=0.9, \beta=0.5.$}}
\label{fig:se4} 
\end{figure}

The above discussions show that the statefinder hierarchy could solve the problems in $Om$ diagnostic for interacting THDE-f and THDE-GO models, however, it meets problem in THDE-H model and the diagnostic results of THDE-GO model with $Q=0$ is not enough satisfactory.
Besides, the single diagnostic method only give us the information from one aspect.
So we consider the supplement method to the statefinder hierarchy, i.e., the composite null diagnostic (CND) in the following.
The CND curves corresponding to $\{S_{3}^{(1)},\epsilon\}$, $\{S_{4}^{(1)},\epsilon\}$ for THDE models including $Q=0$ are plotted in FIGs.~\ref{fig:se3} and \ref{fig:se4}, respectively.
And the fixed points $\{1, 1\}$ of $\Lambda$CDM model are represented by the symbols of star in the figures.
We can see the differences between THDE models and $\Lambda$CDM model by means of CND are slightly distinct because they correspond to the evolving curves and the fixed point, respectively.

From FIGs.~\ref{fig:s3e0a}-\ref{fig:s3e0c} and \ref{fig:s4e0a}-\ref{fig:s4e0c}, it also indicates that the model parameter $\delta$ only influences the values of diagnostic parameters for THDE-H model but it has influence on the trends of curves for THDE-f and THDE-GO models.
And the interaction $Q$ changes the shapes or the trends of the evolutionary trajectories (see FIGs.~\ref{fig:s3eqa}-\ref{fig:s3eqc} and \ref{fig:s4eqa}-\ref{fig:s4eqc}).
Specifically speaking, the interaction $Q$ only influences the shapes of $\{S_{3}^{(1)},\epsilon\}$ curves for THDE-H model but it influences both the shapes and the trends of evolutionary trajectories for THDE-f and THDE-GO models.
As for the curves of $\{S_{4}^{(1)},\epsilon\}$, the interaction $Q$ only changes the shapes of curves for THDE-H and THDE-f models, while they influence both the shapes and the trends for THDE-GO model.
Moreover, it can be obtained that $\epsilon$ evolves towards the direction of increasing $\epsilon$ for THDE-GO model, and it decreases first and then increases for THDE-f and THDE-GO models.
In addition, from the aspect of diagnosing the different values of $\delta$ and different forms of $Q$, it shows that CND performs well for THDE-GO model, and it performs not well for diagnosing the different forms of $Q$ for THDE-H and THDE-f models.
However, it's noted that for THDE-H model, $\{S_{3}^{(1)},\epsilon\}$ and $\{S_{4}^{(1)},\epsilon\}$ breaks the degeneracy among different values of $\delta$ in the future and $\{S_{4}^{(1)},\epsilon\}$ breaks the degeneracy among different forms of $Q$ in the low-redshift and future regions.
It means that CND could provide a great supplement to the statefinder hierarchy.
The present values of the fractional growth parameter $\epsilon_{to}$ and $\triangle \epsilon_{to}=\epsilon_{to}(max)-\epsilon_{to}(min)$ are also given in TABLE \ref{tab:s}.
In addition, we also obtain the information about the growth rate of structure $f(z)$ in the process of calculating $\epsilon(z)$ for THDE models, and we find that the values of $f$ are in line with the observation data $f=0.51\pm0.1$ or $f=0.58\pm0.11$ at the effective redshift \cite{Sen:2005nr}.

\section{Conclusion}
\label{conclusion}
In summary, we have investigated three diagnostic methods, i.e., the $Om$ diagnostic, the statefinder hierarchy $S_{3}^{(1)}$, $S_{4}^{(1)}$ and CND $\{S_{3}^{(1)},\epsilon\}$, $\{S_{4}^{(1)},\epsilon\}$ for three THDE models in order to explore the visualized geometric evolutions corresponding to different values of model parameter and forms of interaction.
We also have obtained the information of the structural growth $f$ and $\epsilon$ through CND, which are important parameters to test a model.

As we discussed above, our results show that model parameter $\delta$ only influences the values of diagnostic parameters for THDE-H model, but it changes the evolutionary trends of THDE-f and THDE-GO models.
Moreover, the evolutionary trends of $S_{3}^{(1)}$, $\{S_{3}^{(1)},\epsilon\}$ could be influenced by the interaction $Q$ for THDE-f and THDE-GO models, while $Q$ only influence the shapes of the curves for THDE-H model.
As for $O_{m}$, $S_{4}^{(1)}$ and $\{S_{4}^{(1)},\epsilon\}$, only the trends of evolutionary trajectories for THDE-GO model could be changed by $Q$.
Thus it can be concluded that the model parameter and forms of interaction could influence the values of diagnostic parameters and evolutionary trends of THDE models.
Furthermore, the figures and tables in this paper have also indicated that combining the statefinder hierarchy $S_{3}^{(1)}$ with CND $\{S_{3}^{(1)},\epsilon\}$ could distinguish the different values of model parameter and different forms of interaction from each other effectively.
It also illustrate that CND could be a proper supplementary method to the statefinder hierarchy and the cosmological information of structure growth has been simultaneously obtained by means of CND.

\section*{Acknowledgments}

This work is supported by the National Natural Science Foundation of China (Grants Nos. 11575075, 11705079 and 11865012).

\end{document}